\documentclass[pdflatex,sn-mathphys-num]{sn-jnl}%

\usepackage{graphicx}%
\usepackage{multirow}%
\usepackage{amsmath,amssymb,amsfonts}%
\usepackage{amsthm}%
\usepackage{mathrsfs}%
\usepackage[title]{appendix}%
\usepackage{xcolor}%
\usepackage{textcomp}%
\usepackage{manyfoot}%
\usepackage{booktabs}%
\usepackage{algorithm}%
\usepackage{algorithmicx}%
\usepackage{algpseudocode}%
\usepackage{listings}%

\usepackage{tikz}
\usetikzlibrary{automata, positioning,shapes,arrows}

\usepackage{booktabs,longtable} %
\usepackage{float, placeins} %
\usepackage{hyperref} %

\usepackage{caption} %
\usepackage{mathtools}
\usepackage{dsfont} %

\theoremstyle{thmstyleone}%

\theoremstyle{thmstyletwo}%
\newtheorem{remark}{Remark}%

\theoremstyle{thmstylethree}%

\begin{document}

\title[Hidden Semi-Markov Models for Fully Functional Time Series]{Hidden Semi-Markov Models for Fully Functional  Time Series}

\author*[1]{\fnm{Jake} \sur{Hoover}}\email{jchoover@ualberta.ca}

\author[1]{\fnm{Adam B} \sur{Kashlak}}\email{kashlak@ualberta.ca}

\author[2]{\fnm{Giseon} \sur{Heo}}\email{gheo@ualberta.ca}

\affil[1]{\orgdiv{Mathematical and Statistical Sciences}, \orgname{University of Alberta}, \orgaddress{\city{Edmonton}, \postcode{T6G 2N8}, \state{AB}, \country{Canada}}}

\affil[2]{\orgdiv{School of Dentistry}, \orgname{University of Alberta}, \orgaddress{\city{Edmonton}, \postcode{T6G 1C9}, \state{AB}, \country{Canada}}}

\abstract{Functional time series have a wide range of practical applications
  to modeling complex datasets 
  such as Electroencephalogram (EEG) and other biological signals.
  Often latent states are hidden within the data dictating how a system
  is behaving and changing over time.  The Hidden Markov Model (HMM) is 
  one common and highly applicable latent state space model where the 
  hidden state sequence evolves as a Markov chain.  While the HMM has 
  had much success modeling discrete, multivariate, and fully functional
  time series, it is also restricted by the Markov property which causes 
  hidden states to persist only at geometrically decaying rates.  This work
  investigates the implementation of hidden semi-Markov models for modeling
  fully functional time series data.  This allows for the hidden state sequence 
  to persist in a given state for a data-driven sojourn time, which can be
  parametrized by a discrete distribution (e.g. negative binomial) or left 
  non-parametric.  For efficient fitting of such a model, the hidden semi-Markov
  chain is embedded into a larger state space Markov chain and classic methods
  like the Baum-Welch and Viterbi algorithms are applied.  We furthermore consider
  functional time series in the general setting of Gaussian measures in locally
  convex topological vector spaces.  This is applied to a time series of 
  pediatric EEG signals collected via overnight polysomnography. 
}



\maketitle

\section{Introduction}

The Hidden Markov Model (HMM) remains one of the most widely used tools for modelling sequential data. Its origins lie in the work of Leonard Baum and his colleagues \citep{BAUM1966, BAUM1970}, which predates the formalization of the expectation-maximization (EM) algorithm \citep{DEMPSTER1977, WU1983}. Originally developed for observations taking values in a finite alphabet, with early applications in speech recognition \citep{FERGUSON1980, RABINER1989, JUANG1991}, the HMM has since been extended to multivariate elliptically symmetric \citep{LIPORACE1982} and skewed \citep{CHATZIS2010} emission distributions, to longitudinal data \citep{ALTMAN2007}, to nonparametric emission densities \citep{DECASTRO2016, GASSIAT2016}, and to functional data \citep{MARTINO2020, SIDROW2021}. Most recently, \citep{KASHLAK_THMM} extended the observation space of the HMM from Euclidean space to general locally convex topological vector spaces, in which each hidden state corresponds to a mean-shifted Gaussian measure. As there is no analogue of Lebesgue measure, and hence no probability density, in infinite-dimensional spaces, their Topological Hidden Markov Model (THMM) replaces the emission density with the Onsager--Machlup functional of the underlying Gaussian measure \citep{ONSAGER1953, BOGACHEV1998}. This allows the analyst to select a Cameron--Martin space suited to the application at hand, and the THMM was shown to successfully cluster sample paths of Brownian motion with drift, fractional Brownian motion, and the Ornstein--Uhlenbeck process, as well as nonparametric curves such as the power spectral densities of electroencephalogram (EEG) signals.

For all of its flexibility in the observation space, the THMM inherits a structural restriction of the classic HMM in its hidden component. Since the hidden state sequence evolves as a homogeneous Markov chain with transition probabilities $a_{ij}$, the number of consecutive time points that the chain spends in state $j$, known as the \emph{dwell time} or \emph{sojourn time} $d_j$, necessarily follows a geometric distribution,
\begin{equation}\label{eq:geomDwell}
  d_j(r) = \mathbb{P}(d_j = r) = a_{jj}^{\,r-1}\,(1 - a_{jj}), \qquad r = 1, 2, \ldots.
\end{equation}
Consequently, regardless of the value of $a_{jj}$, the modal dwell time of every state is a single time step, and the probability of leaving a state does not depend on how long the chain has already occupied it \citep{RABINER1989, LANGROCK}. This memoryless behavior is unrealistic in many applications. 
The THMM was applied to overnight polysomnography data from pediatric patients, the Viterbi path switched between the awake and asleep states considerably more often than the sleep stages labeled by a sleep technician. \cite{KASHLAK_THMM} attributed this behavior to the geometric dwell-time assumption and identified hidden semi-Markov models as a natural remedy. 

Hidden semi-Markov models (HSMMs) remove this restriction by modelling the dwell times explicitly. Upon entering state $j$, the hidden process remains there for a duration drawn from an arbitrary distribution $d_j$ on the positive integers, after which it moves to a new state according to an embedded Markov chain with no self-transitions \citep{FERGUSON1980, LEVINSON1986, YU2010}. The HMM is recovered as the special case in which every $d_j(\cdot)$ is geometric. Originally proposed for speech recognition, HSMMs have since been applied to rainfall \citep{SANSOM2001}, financial returns \citep{BULLA2006}, and animal movement \citep{LANGROCK2012}, among many other areas; see \cite{YU2015} for a comprehensive review. Parameters are typically estimated via an EM algorithm adapted to the semi-Markov setting \citep{GUEDON2003, YU2003}, and implementations are available in \textsf{R} \citep{BULLA2010, OCONNELL2011}. This additional generality is not free, however. The forward--backward recursions of an HSMM must account for every possible duration, so the cost of evaluating the likelihood can grow quadratically in the length of the series in the worst case \citep{GUEDON2003}. Furthermore, as state changes and state durations are modeled separately, the hidden process is no longer Markovian on the original time scale. This complicates the inclusion of covariates, forecasting, and the fitting of stationary models, and most implementations assume that the observed series begins, and possibly ends, at a state boundary.

An elegant alternative, proposed by \cite{LANGROCK} and building on earlier ideas of state aggregation \citep{COX1965, RUSSELL1987, GUEDON2005, JOHNSON2005}, is to partially embed the HSMM as an HMM on an enlarged state space. Each semi-Markov state $j$ is replaced by a \emph{state aggregate} of $m_j$ Markov substates that share a common emission distribution, and the transitions within and out of the aggregate are governed by the discrete hazard rates of the dwell-time probability mass function,
\begin{equation}\label{eq:hazard}
  c_j(r) = \frac{d_j(r)}{1 - D_j(r-1)}, \qquad r = 1, \ldots, m_j,
\end{equation}
where $D_j$ denotes the cumulative distribution function of $d_j$. The dwell-time distribution of the resulting HMM coincides with $d_j$ for durations of up to $m_j$ time steps and has a geometric tail thereafter. Hence, the representation is exact for dwell-time distributions with finite support or with geometric tails, and it can be made arbitrarily accurate for all others by enlarging $m_j$. The expanded model has exactly as many parameters as the HSMM it represents, its likelihood can be evaluated in time linear in the length of the series, and, being an HMM, it inherits the full toolkit of HMM methodology---forward--backward recursions, Viterbi decoding, stationary initial distributions, and state prediction---without modification.

In this article, we unite these two lines of research and propose the \emph{Topological Hidden Semi-Markov Model} (THSMM). As in the THMM, observations may take values in a general locally convex space, encompassing finite-dimensional data, functional data, and sample paths of stochastic processes among others. Each hidden state corresponds to a mean-shifted Gaussian measure, and emissions are evaluated through the Onsager--Machlup functional. Unlike the THMM, each hidden state is additionally endowed with its own dwell-time distribution. These may be drawn from parametric families such as the (shifted) Poisson, (shifted) negative binomial, or discretized Weibull distributions, or they may be left nonparametric, either on a finite support or with an unstructured start and a geometric tail in the spirit of \cite{SANSOM2001}. The semi-Markov structure is implemented via the state-aggregate construction of \cite{LANGROCK}.

A key feature of this construction is that the two extensions decouple cleanly. Since every substate in an aggregate shares the same Onsager--Machlup emission functional, the topological component of the model enters only through the emission terms, whereas the semi-Markov component enters only through the structured transition matrix. As a result, the posterior probability of occupying a given semi-Markov state is simply the sum of the posterior probabilities of its substates, and the state means are re-estimated by the same weighted criteria as in the THMM, with these aggregate posterior probabilities serving as the weights. The THMM is recovered when every dwell-time distribution is geometric, and the classic HSMM with Gaussian emissions sharing a common covariance matrix is recovered when the observation space is $\mathbb{R}^d$.

To our knowledge, explicit modeling of state durations has received little attention outside of finite-dimensional observation spaces. Existing HMMs for functional data, such as \cite{MARTINO2020, SIDROW2021}, do not model the state durations explicitly, while existing HSMM methodology, including Bayesian nonparametric variants \citep{JOHNSON2013}, assumes emission distributions that admit densities with respect to Lebesgue or counting measure.

The THSMM fills this gap, allowing realistic sojourn behaviour to be modelled jointly with emissions living in infinite-dimensional spaces. We demonstrate the THSMM on simulated diffusion processes whose hidden states have non-geometric dwell times, and we revisit the pediatric obstructive sleep apnea polysomnography data of \cite{KASHLAK_THMM} to show how explicit dwell-time modelling affects the predicted sequence of sleep states.

The remainder of this article is organized as follows. Section \ref{sec:hsmmFctData} outlines the general theory for using HSMMs with functional data. Section \ref{sec:hmmEmbedding} reviews \cite{LANGROCK} and proposes an EM framework for model fitting. Section \ref{sec:sim} explores various simulated data sets to give a comprehensive overview of the models capabilities. Section \ref{sec:osaData} applies the THSMM to the identification of sleep states form pediatric EEG data. For pseudo code of algorithms and in depth derivations please reference Appendices \ref{apx:hmmEmEquations} and \ref{apx:hmmEmPseudoCode}. 

\section{HSMM for functional data}\label{sec:hsmmFctData}

In this section we define a HSMM that utilizes functional data (THSMM); even though this construction is quickly dissembled in the next section it is important to fully establish it. For an in depth study of the HSMM, extensions, and case studies, we refer the reader to \cite{YU2015}. 

\begin{figure}[ht]
    \centering
    \resizebox{\textwidth}{!}{%
        \begin{tikzpicture}[
            > = stealth,
            shorten > = 1pt,
            auto,
            node distance = 3cm,
            semithick
        ]
        \tikzset{
            every state/.style={
                draw=black,
                thick,
                fill=white,
                minimum size=4mm
            },
            no circle/.style={
                draw=none,
                minimum size=0mm,
                inner sep=0mm
            }
        }

        \node[state] (S1) {$S_1 = 2$};
        \node[no circle] (X1) [above of=S1] {$2 \ 2 \ 2$};
        \node[state] (S2) [right of=S1] {$S_2 = 3$};
        \node[no circle] (X2) [above of=S2] {$3 \ 3$};
        \node[state] (S3) [right of=S2] {$S_3 = 2$};
        \node[no circle] (X3) [above of=S3] {$2$};
        \node[state] (S4) [right of=S3] {$S_4 = 3$};
        \node[no circle] (X4) [above of=S4] {$3 \ 3 \ 3$};
        \node[state] (S5) [right of=S4] {$S_5 = 1$};
        \node[no circle] (X5) [above of=S5] {$1 \ 1 \ 1 \ 1$};
        \node[no circle] (Blank) [right of=S5] {$\dots$};

        \path[->] (S1) edge node[left] {$d_2(3)$} (X1);
        \path[->] (S2) edge node[left] {$d_3(2)$} (X2);
        \path[->] (S3) edge node[left] {$d_2(1)$} (X3);
        \path[->] (S4) edge node[left] {$d_3(3)$} (X4);
        \path[->] (S5) edge node[left] {$d_1(4)$} (X5);

        \path[->] (S1) edge (S2);
        \path[->] (S2) edge (S3);
        \path[->] (S3) edge (S4);
        \path[->] (S4) edge (S5);
        \path[->] (S5) edge (Blank);
        \end{tikzpicture}
    } %
    \caption{An example of a semi-Markov process where the state process is determined by transition probability matrix $\Omega$ and stationary distribution $\pi$ and the duration of the stay is determined by $d_i(r)$.}
    \label{fig:semiMarkovProcess}
\end{figure}
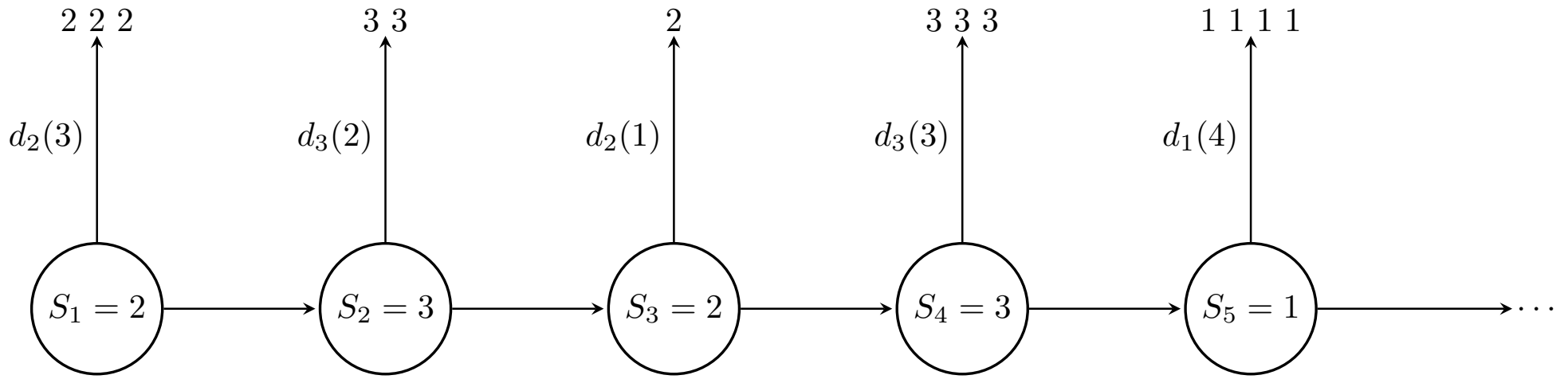

Let $O_1,\ldots,O_T$ be a sequence of observations taking values in a locally
convex topological vector space $\mathcal{X}$, and let $S_1,\ldots,S_T$ be an unobserved state sequence taking values in $\{1,\ldots,N\}$. As in the THMM \citep{KASHLAK_THMM}, the observations are assumed conditionally independent given the hidden states, and the conditional law of $O_t$ given $S_t=i$ is a Gaussian measure on $\mathcal{X}$ with state-specific mean $h_i$ and a
shared covariance structure. Unlike the THMM, the hidden process $(S_t)$ is not assumed to be a Markov chain on the original time scale; instead it is a semi-Markov chain whose dwell times are modelled explicitly, see Figure \ref{fig:semiMarkovProcess}. We describe the hidden component in Section~\ref{sec:thsmm:hidden}, the emission component in Section~\ref{sec:thsmm:emission}, and the resulting likelihood and its HMM representation in Sections~\ref{sec:hmmEmbedding}.

\subsection{The Hidden Semi-Markov Chain}\label{sec:thsmm:hidden}
 
The hidden process is described by two components. The first is an
\emph{embedded} Markov chain governing which state is visited next, with
transition probability matrix (tpm) $\Omega = (\omega_{ij})_{i,j=1}^N$,
\begin{equation}\label{eq:omega}
  \omega_{ij} = \mathbb{P}\left(S_{t+1} = j \mid S_t = i\right),
  \qquad \omega_{ii} = 0,\qquad \sum_{j\neq i}\omega_{ij} = 1.
\end{equation}
That is, $\omega_{ij}$ is the probability of moving to state $j$ given that the
process leaves state $i$; self-transitions are excluded, since the time spent in a
state is governed entirely by the second component. We assume $\Omega$ is
irreducible and denote its stationary distribution by $\pi$; i.e.
$\pi^\top\Omega = \pi^\top$ and $\pi^\top\mathbf{1} = 1$. Note that $\pi_i$ is
the long-run proportion of \emph{visits} to state $i$, not the long-run proportion
of \emph{time} spent there; the two are related in
Remark~\ref{rem:stationary}.
 
The second component is a collection of \emph{dwell-time} (or sojourn) distributions. Let 
$d_j(r) = \mathbb{P}(\text{dwell time in state } j = r)$, $r \in \mathbb{N}$, 
denote the probability mass function of the dwell time in state $j$, and $D_j(r) = \sum_{s=1}^{r} d_j(s)$ its cumulative distribution function. Upon entering state $i$, the process remains there for a random number of time steps $r$ with probability $d_i(r)$ after which it moves to a new state $j\neq i$ with probability $\omega_{ij}$. We write
\begin{equation}\label{eq:survHaz}
  \bar D_i(r) = \mathbb{P}(D_i \geq r) = \sum_{u\geq r} d_i(u),
  \qquad
  c_i(r) = \mathbb{P}(D_i = r \mid D_i \geq r) = \frac{d_i(r)}{\bar D_i(r)},
\end{equation}
for the survival function and the discrete hazard rate of $D_i$, with the
convention $c_i(r) = 1$ whenever $\bar D_i(r) = 0$. The hazard $c_i(r)$ is the probability that a sojourn in state $i$ ends after exactly $r$ steps, given that it has lasted at least $r$ steps, in this sense it can be viewed as a sort of right censoring. Notice that this definition coincides with the quantity $d_j(r)/\{1-D_j(r-1)\}$ of Equation~\eqref{eq:hazard}. Dwell times of successive visits are independent of one another and of the embedded chain.

The dwell-time distributions may be drawn from any family on the positive integers. Those considered in this article---the geometric, (shifted) Poisson, 
(shifted) negative binomial, discretized Weibull, and nonparametric distributions with or without a geometric tail---are detailed in Section~\ref{sec:hmmEmbedding}. In the special case $d_i(r) = (1-p_i)^{r-1}p_i$, the hazard $c_i(r) \equiv p_i$ is constant, the process is memoryless, and $(S_t)$ is a homogeneous Markov chain with transition probabilities
\begin{equation}\label{eq:geomReduction}
  a_{ii} = 1 - p_i, \qquad a_{ij} = p_i\,\omega_{ij}\quad (j\neq i).
\end{equation}
 
\subsection{Emissions via the Onsager--Machlup functional}\label{sec:thsmm:emission}
 
The emission component is inherited unchanged from the THMM; we summarize it
here and refer to \cite{KASHLAK_THMM} for details. 
Let $\gamma$ be a centred Radon Gaussian measure on $\mathcal{X}$ with covariance operator $R_\gamma$ and Cameron--Martin space $H = H(\gamma)$, equipped with norm $|\cdot|_H$. Conditional on $S_t = i$, the observation is modelled as
\begin{equation}\label{eq:emissionModel}
  O_t = h_i + Z_t, \qquad Z_t \overset{\text{iid}}{\sim} \gamma,
  \qquad h_i \in H,
\end{equation}
so that each hidden state corresponds to a shift of the common Gaussian
measure $\gamma$ by a state mean $h_i$ in an appropriate
Cameron--Martin space. Since
$\mathcal{X}$ is in general infinite-dimensional, the law of $O_t$ has no
density with respect to a sigma-finite translation invariant measure
(e.g. Lebesgue measure), and we instead use the
Onsager--Machlup functional of $\gamma$ as the emission function,
\begin{equation}\label{eq:OM}
  b_i(O_t)
  = \lim_{\varepsilon\to 0}
    \frac{\gamma\left(V_\varepsilon + \{O_t - h_i\}\right)}{\gamma(V_\varepsilon)}
  = \exp\left( -\tfrac{1}{2}\left|\Pi_q\{O_t - h_i\}\right|_H^2 \right),
\end{equation}
where $V_\varepsilon = \{x\in\mathcal{X} : q(x)\leq\varepsilon\}$ is the $\varepsilon$-ball of a seminorm $q$ on $\mathcal{X}$, and $\Pi_q$ is the orthogonal projection in $H$ onto the orthogonal complement of $\{a\in H : q(a) = 0\}$ \citep[Corollary~4.7.8]{BOGACHEV1998}. When $q$ is a norm, $\Pi_q$ is the identity and \eqref{eq:OM} reduces to
$b_i(O_t) = \exp(-\tfrac12|O_t - h_i|_H^2)$.

The choice of $\gamma$, and hence of $H$, is left to the analyst. For example, when $\mathcal{X} = \mathbb{R}^d$ and $\gamma = \mathcal{N}(0,\Sigma)$, we have $|h|_H^2 = h^\top\Sigma^{-1}h$ and $b_i$ is, up to a constant, the Gaussian density with mean $h_i$ and common covariance $\Sigma$. When $\gamma$ is the Wiener measure on $C_0[0,1]$, $H = W_0^{2,1}[0,1]$ and
\[
  b_i(O_t) = \exp\left( -\frac12 \int_0^1 \left| \dot O_t(\tau) - \dot h_i(\tau) \right|^2 d\tau \right),
\]
and analogous expressions are available for Brownian motion with drift, the Ornstein--Uhlenbeck process, and fractional Brownian motion \cite{KASHLAK_THMM} [Section~4].
 
\begin{remark}[Why the Onsager--Machlup functional suffices]\label{rem:OM} Fix $\varepsilon > 0$ and replace each emission density in the HSMM likelihood by the small-ball probability $\gamma(V_\varepsilon + \{O_t - h_{s_t}\})$. Every time point contributes exactly one emission factor, regardless of how the sequence $s_1,\ldots,s_T$ is partitioned into sojourns, so dividing this small-ball likelihood by $\gamma(V_\varepsilon)^T$---a constant free of all model parameters---and letting $\varepsilon\to 0$ term by term in the finite sum over state sequences yields the analogue of the likelihood in Equation \eqref{eq:likHSMM} below, in which each emission is the Onsager--Machlup functional \eqref{eq:OM}. Hence the justification of \cite{KASHLAK_THMM} carries over to the semi-Markov setting without modification: the semi-Markov structure alters only the weights placed on each state sequence, not the emission terms.
\end{remark}

\section{HMM Embedding}\label{sec:hmmEmbedding}

The work of \cite{LANGROCK} consider specially structured HMMs that capture --- up to an approximation --- the sojourn distribution component of a HSMM. Below we give a brief summary of the technique, the advantages and disadvantages associated, and an extension for the use of the Baum-Welch algorithm for parameter estimation as opposed to the numerical maximization approach used in the original work. 

\subsection{HMMs with Arbitrary State Dwell-Time Distributions}

We begin by outlining the technique. Let the N-state HSMM be represented by $\Psi = (\Omega,d_i(r),b_i(x_t))$, a state dependent process $X_{1:T}$ with realizations $x_{1:T}$ and Markov chain $S_{1:T}$ with realizations $j \in \{1,\dots,N\}$ prescribing to tpm $\Omega$. For each state $j$ define the size of a state aggregate $m_i \in \mathbb{N}^{>0}, \ i=1,\dots N$ with $m_0 =0$. Consider the HMM with state dependent process $X_{1:T}^\star$, Markov chain $S_{1:T}^\star$ with states $\{1,2,\dots,\sum_{i=1}^N m_i\}$. Refer to the sets
\begin{align*}
    I_k = \left\{ n | \sum_{i=0}^{k-1} m_i < n \leq \sum_{i=0}^k m_i \right\}
\end{align*}
as state aggregates and define $i_k^- \coloneqq \min(I_k)$ and $i_k^+ \coloneqq \max(I_k)$. The state aggregate $I_k$ is the index set corresponding to HMM states associated with state $k$ of the HSMM. Assume that HMM states in aggregate $I_k$ are associated with the same emission distribution $b_k(\cdot)$, that is 
\begin{align*}
    \mathbb{P}(X_t^\star|S_t^\star \in I_k) = \mathbb{P}(X_t|S_t=k)\text{.}
\end{align*}

Define the hazard rate function $c_k(r)$ by:
\begin{align*}
    c_k(r) \coloneqq \begin{cases}
        \frac{d_k(r)}{1 - D_k(r-1)} & \text{ for } D_k(r-1) < 1, \\
        1  & \text{ for } D_k(r-1)=1.
    \end{cases}
\end{align*}
Where $D_k(r)=\sum_{s\le r}d_k(s)$, and $\bar D_k(r)=1-D_k(r-1)=P(\text{dwell}\ge r)$ the censoring factor.
The hazard rate is how the sojourn distribution is embedded in the HMM construction. 

Denote the HMM tpm by $\Gamma$ and construct it as follows. Build a $N\times N$ block matrix (\ref{eq:gammaHmmExpanded}) whose $(i,j)$th block is size $m_i \times m_j$. We populate the blocks (\ref{eq:diagBlock}) and (\ref{eq:offDiagBlock}) using the hazard rates in such a way that the sojourn distribution is exactly embedded in the HMMs Markov chain up to a duration less than or equal to its respective aggregate size.

\begin{align}\label{eq:gammaHmmExpanded}
    \mathbf{\Gamma} = 
    \begin{bmatrix}
        \mathbf{\Gamma}_{11} & & \hdots & & \mathbf{\Gamma}_{1m} \\
         & & & & \\
        \vdots & & \ddots & & \vdots \\
         & & & & \\
         \mathbf{\Gamma}_{m1} & & \hdots & & \mathbf{\Gamma}_{mm}
    \end{bmatrix}.
\end{align}
The $m_i \times m_i$ diagonal block $\mathbf{\Gamma}_{ii}$, $i=1,\dots,m$ are defined, for $m_i \geq 2$, as 
\begin{align}\label{eq:diagBlock}
    \mathbf{\Gamma_{ii}} =
    \renewcommand{\arraystretch}{1.4}
    \setlength{\arraycolsep}{7pt}
    \left[
    \begin{array}{c|cccccccc}
        0 & 1 - c_i(1) & 0          & \cdots     &            &        &        &        & 0 \\
        0 & 0          & 1 - c_i(2) & 0          & \cdots     &        &        &        & 0 \\
        0 &            &            & \ddots     &            &        &        &        & 0 \\
        0 &            &            &            & 1 - c_i(r) &        &        &        & 0 \\
        0 &            &            &            &            & \ddots &        &        & 0 \\
        0 &            &            &            &            &        &        &        & 1 - c_i(m_i - 1) \\ \hline
        0 &            &            &            &            &        &        &        & 1 - c_i(m_i)
    \end{array}
    \right],
\end{align}
where the entries $1-c_i(1)$ to $1-c_i(m_i-1)$ are along the super diagonal. The $m_i \times m_j$ off diagonal blocks $\mathbf{\Gamma}_{ij}$ are defined as
\begin{align}\label{eq:offDiagBlock}
{\mathbf\Gamma_{ij}}=
    \begin{bmatrix}
        c_i(1) \omega_{ij} & 0 & \dots & 0 \\
        c_i(2) \omega_{ij} & 0 & \dots & 0 \\
        \vdots & \vdots & & \vdots  \\
        c_i(m_i) \omega_{ij} & 0 & \dots & 0 \\
    \end{bmatrix}.
\end{align}

There are two important consequences of this construction. Firstly the state transitions between aggregates in the HMM align with the state transitions in the HSMM. Specifically, for $i \neq j$, $1\leq i,j \leq N$, $\mathbb{P}(S_{t+1}^\star \in I_j|S_t^\star \in I_i, S_{t+1}^\star \notin I_i) = \mathbb{P}(S_{t+1}=j|S_t = i, S_{t+1}\neq i)$. Furthermore, the duration with state aggregate $k$ corresponds to the sojourn distribution for state $k$ up to a duration $m_i$ and has geometric decay thereafter. That is, for any $k \in \{1,\dots,N\}$ 
\begin{align*}
    d_k^\star(r) =  \begin{cases}
        d_k(r) & \text{ for } r \leq m_i,\\
        d_k(m_k)(1-c_k(m_k))^{r-m_k} & \text{ for } r > m_k.
    \end{cases}
\end{align*}
The geometric decay is a direct consequence of Markov chain behavior. One can see that the difference between $d_k^\star$ and $d_k$ can be made arbitrarily small by choosing $m_i$ sufficiently large. Also, any sojourn distribution with finite support can be exactly recreated by choosing $m_k$ to be the maximum dwell time in state $k$ having a non-zero probability.
 
Let $\Theta:=(\Omega,d_1,\ldots,d_N,h_1,\ldots,h_N, \delta)$ be the collection of model parameters. 
A state sequence
$s = (s_1,\ldots,s_T)$ may be written uniquely as a sequence of $M$ sojourns
$(j_1,r_1),\ldots,(j_M,r_M)$, where $j_m$ is the state visited in the $m$-th
sojourn, $r_m$ its length, $j_{m+1}\neq j_m$, and $r_1+\cdots+r_M = T$. If the
series is assumed to begin at the start of a sojourn, with
$\mathbb{P}(S_1 = i) = \delta_i$, the probability of $s$ is
\begin{equation}\label{eq:pathProb}
  \mathbb{P}(s)
  = \delta_{j_1}
    \left\{ \prod_{m=1}^{M-1} d_{j_m}(r_m)\, \omega_{j_m j_{m+1}} \right\}
    \bar D_{j_M}(r_M),
\end{equation}
where the survival function in the final factor accounts for the last sojourn
being right-censored at time $T$. The log likelihood of the THSMM is then
\begin{equation}\label{eq:likHSMM}
  l(\Theta \mid X_{1,\ldots,T})
  = \log \sum_{s\in\{1,\ldots,N\}^T} \mathbb{P}(s) \prod_{t=1}^T b_{s_t}(X_t).
\end{equation}
Direct evaluation of \eqref{eq:likHSMM} through the semi-Markov forward recursion
requires summing over all possible sojourn lengths at each time step, at a cost
that grows quadratically in $T$ in the worst case \citep{GUEDON2003}. Moreover,
\eqref{eq:pathProb} requires the series to begin at a sojourn boundary; relaxing
this is awkward because $(S_t)$ is not Markov on the original time scale. Both
issues are resolved by the state-aggregate representation.

\begin{remark}[Stationarity]\label{rem:stationary}
Let $\mu_i^* = \sum_{r\geq 1} r\,d_i^*(r)$ denote the mean dwell time in aggregate $I_i$, and suppose $c_i(K_i) > 0$ for every $i$. The stationary distribution $\delta^*$ of $\Gamma$ is
\[
  \delta^*_{(i,r)} \propto
  \begin{cases}
    \pi_i\, \bar D_i(r), & r < K_i, \\[2pt]
    \pi_i\, \bar D_i(K_i) / c_i(K_i), & r = K_i,
  \end{cases}
\]
normalized to sum to one. Summing over each aggregate, the stationary probability
of hidden state $i$ is
\[
  \mathbb{P}(S_t = i) = \sum_{r=1}^{K_i} \delta^*_{(i,r)}
  = \frac{\pi_i\, \mu_i^*}{\sum_{k=1}^N \pi_k\, \mu_k^*},
\]
so that the long-run proportion of time spent in state $i$ is the proportion of visits to state $i$ under the embedded chain, reweighted by the mean dwell time of state $i$.
\end{remark}

The usefulness of such a construction is that by porting a HSMM into a HMM the user immediately gains access to a wide array of well established HMM theory. Of particular interest is that it greatly simplifies the likelihood of a HSMM and in turn the maximization of such a likelihood. One now has a greatly simplified modular framework that supports a wide range of sojourn distributions without the need for custom infrastructure. This is well demonstrated in \cite{LANGROCK} where a maximization of a HSMM over many different sojourn distributions can be done with a simple numerical maximization of the constructed HMM likelihood without the need for significant tailoring of the algorithm. One simply needs to build the hazard rate equations for any specific sojourn they wish to use. 

The expanded state space does not increase the size of the parameter 
space as the HMM formulation of $\Gamma$ a deterministic function of $\Omega$ and the dwell parameters from the HSMM and therefore there are as many constrained parameters in both models. 
Computational slowdown in this expanded state space can be mitigated
via sparse matrix
computations (in \textrm{C++} or in R using package \textrm{TMB} \citep{TMB})
yielding
speedups in runtime over the naive implementation.

\subsection{Model Fitting Via Baum-Welch (EM algorithm):}\label{sec:modelFitting}

Unlike in the original work of \cite{LANGROCK} we fit the HSMM (embedded into a HMM) via the Baum-Welch algorithm. 
This is for two reasons: the first being that direct numerical maximization 
of the parameters for a measure in an infinite dimensional space 
is often ill-posed whereas the M-step in Baum-Welch provides 
a closed form solution. 
Additionally, we can derive closed for updates for the hazard rates and in turn the tpm $\Omega$, and weighted least square estimates (WLSE) for sojourn distributions. Throughout the optimization there is no need to construct $\Gamma$ circumventing the issue of doing computations with large matrices. 

In the proceeding discussion we introduce the following notation to describe state transitions in $\Gamma$. Define $\Gamma((i,l)\rightarrow(j,k))$ to be the probability of transitioning from the $l$-th entry of state aggregate $i$ to the $k$-th entry of state aggregate $j$. There are only two such transitions, advance within a state aggregate $i$ with probability:
\begin{align*}
    (j,s)\to(j,s^+),\qquad s^+=\min(s+1,\,m_j),\qquad \text{probability } 1-c_j(s),
\end{align*}
or advance to a new aggregate $j$ with probability 
\begin{align*}
    \Gamma((i,r)\rightarrow(j,1)) = \omega_{i,j}c_i(r), \quad j \neq i.
\end{align*}
All remaining entries of $\Gamma$ being equal to zero. Note that we need the construction $s^+$ to appropriately deal with the geometric tail of a state aggregate.

The forward ($\alpha$) and backwards ($\beta$) equations handle the expectation step of the Baum-Welch algorithm and are defined as follows:
\begin{align*}
    \alpha_t(i,r) &= \mathbb{P}(x_{1:t}, S_t=(i,r)) \\
    \beta_t(i,r) &= \mathbb{P}(x_{t+1:T}|S_t=(i,r))
\end{align*}
Before calculating these probabilities it is useful to define the following:
\begin{align}
    F_t(i) &= \sum_{r=1}^{m_i} \alpha_t(i,r)c_i(r), \label{eq:F} \\
    G_t(i) &= \sum_{i \neq j} \omega_{i,j} b_{t+1}(j,1) \beta_{t+1}(j,1). \label{eq:G}
\end{align}
Where Equation \ref{eq:F} can be seen as the total probability mass leaving aggregate $i$ at time $t$ weighted by the hazard and summed over the substates, and Equation \ref{eq:G} can be seen as the probability of all the data after time $t$ given that state $i$ ends at time $t$. We then formulate the recursions for the $\alpha$ and $\beta$ passes as follows: 
\begin{align}\label{eq:hmmEmAlpha}
       \alpha_{t+1}(j,l) &= b_{t+1}(j,l)\Big[\sum_{s:\,s^+=l}\alpha_t(j,s)\{1-c_j(s)\}
     + \mathbf{1}\{l=1\}\sum_{i\neq j}\omega_{i,j}F_t(i)\Big] \\
    \label{eq:hmmEmBeta}
    \beta_t(i,r) &=\{1-c_i(r)\}\,b_{t+1}(i,r^+)\,\beta_{t+1}(i,r^+)+c_i(r)\,G_t(i).
\end{align}
Case by case we have the following situations for $\alpha_{t+1}(j,l)$:
\begin{align*}
\alpha_{t+1}(j,1) &= b_{t+1}(j,1)\Big[\sum_{i\neq j}\omega_{i,j}F_t(i)+\mathbf 1\{m_j=1\}\,\alpha_t(j,1)\{1-c_j(1)\}\Big],\\
\alpha_{t+1}(j,l) &= b_{t+1}(j,l)\,\alpha_t(j,l-1)\{1-c_j(l-1)\}, \qquad 2\le l<m_j,\\
\alpha_{t+1}(j,m_j) &= b_{t+1}(j,m_j)\Big[\alpha_t(j,m_j-1)\{1-c_j(m_j-1)\}+\alpha_t(j,m_j)\{1-c_j(m_j)\}\Big], \qquad m_j\ge2.
\end{align*}
The first form of each pass handles the probability of advancing within a state aggregate and the second term handles the probability of leaving an aggregate. Using the Equations \ref{eq:hmmEmAlpha} and \ref{eq:hmmEmBeta}, we can iteratively produce the necessary posterior probabilities that act as weights in the maximization step. 

To derive estimates for the hazard rates and the HSMM tpm entries we need to keep track of two things, how often the chain remained in a state and how often the chain left a state. We define the following counts:
\begin{align*}
    \eta_i(r) &= \frac{1}{\mathbb{P}(x_{1:T})} \sum_{t=1}^{T-1} \alpha_t(i,r) \beta_t(i,r),\\
    \Delta_i(r) &= \frac{c_i(r)}{\mathbb{P}(x_{1:T})} \sum_{t=1}^{T-1} \alpha_t(i,r)G_t(i),\\
    N(i,j) &= \frac{1}{\mathbb{P}(x_{1:T})} \sum_{t=1}^{T-1} \omega_{i,j} b_{t+1}(j,1) \beta_{t+1}(j,1) F_t(i).
\end{align*}
Where $\eta_i(r)$ represents the number of visits to state $(i,r)$, $\Delta_i(r)$ represents the number of departures from state $i$ that happened at duration $r$, and $N(i,j)$ the total number of jumps from state $i$ to state $j$. In addition, let the smoothed posterior probabilities be
\begin{align*}
    \iota_t(i,r) = 
    \frac{\alpha_t(i,r)\beta_t(i,r)}{\mathbb{P}(x_{1:T})} 
    \text{ and } 
    \iota_t(i) = \sum_{r=1}^{m_i} \iota_t(i,r) \ =\mathbb{P}(S_t \in i|x_{1:T})\footnotemark.
\end{align*}
\footnotetext{Yes we are starting to run out of symbols.}
Denote the full log likelihood of the HMM by $\log P(S,x)$:
\begin{align*}
\log P(S, x) &= \sum_i \sum_r \mathbf{1}\!\left[S_1 = (i,r)\right] \log \delta(i,r) \\
&\quad + \sum_{i} \sum_{j \neq i} \left[ \sum_{t=1}^{T-1} \sum_{r} \mathbf{1}\!\left[S_t = (i,r),\, S_{t+1} = (j,1)\right] \right] \log \omega(i,j) \\
&\quad + \sum_{i} \sum_{r=1}^{m_i} \Bigg\{ \left[ \sum_{t=1}^{T-1} \sum_{j \neq i} \mathbf{1}\!\left[S_t = (i,r),\, S_{t+1} = (j,1)\right] \right] \log c_i(r) \\
&\quad+ \left[ \sum_{t=1}^{T-1} \mathbf{1}\!\left[S_t = (i,r),\, \text{stay}\right] \right] \log\!\big(1 - c_i(r)\big) \Bigg\} \\
&\quad + \sum_{i} \sum_{t=1}^{T} \left[ \sum_{r} \mathbf{1}\!\left[S_t = (i,r)\right] \right] \log f(x_t; \theta_i)
\end{align*}

We then define $Q$ as the expected value of the log likelihood $Q(\Theta\mid\Theta^{(k)})=\mathrm{E}_{\Theta^{(k)}}\big[\log P(S,x;\Theta)\mid x_{1:T}\big]$:
\begin{align*}
Q &= \underbrace{\sum_i \sum_r \iota_1(i,r) \log \delta(i,r)}_{Q_{\text{init}}}
+ \underbrace{\sum_{i} \sum_{j \neq i} N(i,j) \log w(i,j)}_{Q_w} \\
&\quad + \underbrace{\sum_{i} \sum_{r=1}^{m_i} \Big[ \Delta_i(r) \log c_i(r) + \big(\eta_i(r) - \Delta_i(r)\big) \log\!\big(1 - c_i(r)\big) \Big]}_{Q_{\text{dwell}}} \\
&\quad + \underbrace{\sum_{i} \sum_{t=1}^{T} \iota_t(i) \log b(x_t; \theta_i)}_{Q_{\text{emit}}}
\end{align*}
Estimates are then found by maximizing the specific summand of Q corresponding to those respective parameters and any constraints surrounding them. 

HSMM transition probabilities $\omega_{i,j}$ are updated by maximizing $Q_\omega$ subject to the constraint $\sum_{j\neq i}\omega_{i,j} = 1$ and are calculated as follows:
\begin{equation}
    \hat{\omega}_{i,j} = \frac{N(i,j)}{\sum_{l \neq i}N(i,l)}.
\end{equation}

One can then either find the stationary distribution of the HSMM or update the initial distribution for the HMM via the EM. Assuming the initial distribution puts mass only on the first substate of each state, that is $\delta(i,1)=\delta_i$ and $\delta(i,r) = 0$ for $r > 1$, we can rewrite $Q_{init}$ as:
\begin{align*}
    Q_{init} = \sum_i \iota_1(i,1) \log(\delta_i).
\end{align*}
Which can be maximized subject to the constrain that $\sum_i \delta_i = 1$, yielding:
\begin{align*}
    \hat\delta_i=\iota_1(i,1)=\alpha_1(i,1)\beta_1(i,1)/\mathbb P(x_{1:T})
\end{align*}

Emission functions are updated the same as in \cite{KASHLAK_THMM} and we refer to the reader there for a detailed summary of derivations. A condensed update schema can be found in \ref{apx:hmmEmPseudoCode} for each functional emission satisfying:
\begin{align*}
    \hat{\theta}_i = \arg \max_\theta \sum_{t=1}^T \iota_t(i) \log f(x_t;\theta).
\end{align*}

The final component is to maximize the sojourn parameters. Recall the component of $Q$ that corresponds to sojourn distributions:
\begin{equation*}
    Q_{dwell} = \sum_{i} \sum_{r=1}^{m_i} \Big[ \Delta_i(r) \log c_i(r) + \big(\eta_i(r) - \Delta_i(r)\big) \log\!\big(1 - c_i(r)\big) \Big].
\end{equation*}
At each duration $r$ there are $\eta_i(r)$ ``trials'' and $d_i(r)$ ``successes'' with the probability of success (a departure) being $c_i(r)$. In this view this is the log likelihood of a Binomial distribution. We can then develop update equations for families of sojourn distributions.

\textbf{Non-Parametric (NP) Sojourn Distribution Updates:} 
Each $c_i(r)$ is a free parameter that can be maximized individually. We start by finding the hazard rate update:
\begin{equation*}
    \hat{c}_i(r) = {\Delta_i(r)}/{\eta_i(r)},
\end{equation*}
then convert back to probabilities:
\begin{equation}
    \hat{p}_i(r) = \hat{c}_i(r) \prod_{s<r}(1-\hat{c}_i(s)).
\end{equation}
When the NP sojourn has no tail, the process must leave the state at
or before duration $m_i$.  Thus,
\begin{equation*}
    c_i(m_i) = 1 \Rightarrow \hat{c}_i(m_i)=1,
\end{equation*}
yielding the desired property of no tail. When the NP sojourn has a tail, the decay of staying within the state is a geometric distribution, and we have:
\begin{equation*}
    \hat{c}_i(m_i) = {\Delta_i(m_i)}/{\eta_i(m_i)}.
\end{equation*}
If $c_i(r) = 0$, then the Markov chain has an absorbing state in aggregate $i$ duration $r+1$. For numerical stability, it is best to initialize the parameters away form zero as the EM algorithm will not be able to move it.  

\textbf{Geometric Sojourn Distribution Updates:} 
A single hazard $c_i(r) = p_i$ governs the probabilities of transitions. Thus we have:
\begin{equation}
    \hat{c}_i(r) = {\sum_r \Delta_i(r)}/{\sum_r \eta_i(r)}.
\end{equation}

\textbf{Parametric Sojourn Distribution Updates:} 
Here we encompass various families such as Poisson, Negative Binomial, and Weibull. In the case of a sojourn distribution that follows one of the preceding distributions, the hazard rate is a curve 
$c_i(r) = h(r;\phi_i)$ or $h_{r,i}$ for
short parameterized by a vector $\phi_i$. This $Q_{dwell}$ becomes:
\begin{equation*}
    Q_{dwell} = \sum_{i} \sum_{r=1}^{m_i} \Big[ \Delta_i(r) \log h(r;\phi_i) + \big(\eta_i(r) - \Delta_i(r)\big) \log\!\big(1 - h(r;\phi_i)\big) \Big].
\end{equation*}
We can now update each $\phi_i$ 
via Iteratively Reweighted Least Squares (IRLS). 
Differentiating a single term of $Q_{dwell,i}$ with respect to $\phi_i$
gives:
\begin{equation*}
    \frac{d}{d \phi_i}Q_{dwell,i} = 
    \left[
      \frac{\Delta_i(r)}{h_{r,i}} + \frac{\eta_i(r) - \Delta_i(r)}{1-h_{r,i}}
    \right] \dot{h}_{r,i},
\end{equation*}
and
summing over $r$ gives the score:
\begin{equation*}
    U(\phi_i) = \sum_{r=1}^{m_i} 
    \frac{\Delta_i(r) - \eta_i(r)h_{r,i}}{h_{r,i}(1-h_{r,i})} \dot{h}_{r,i}.
\end{equation*}
Using the reasoning above and treating $\Delta_i(r)$ as a Binomial$(n = \eta_i(r), p = c_i(r))$, we can compute the Hessian and the resulting Fisher information 
as follows.
\begin{align*}
    \frac{d^2}{d \phi^2}Q_{dwell,i} &= 
    \left[\frac{-d_r}{h_{r,i}^2} - 
    \frac{\eta_i - \Delta_i}{(1 - h_{r,i})^2}\right] \dot{h}_{r,i} \dot{h}_{r,i}^T
    + \left[\frac{d_r}{h_{r,i}} - \frac{\eta_i - \Delta_i}{1 - h_{r,i}}\right] \ddot{h}_{r,i}.
\end{align*}
Summing over $r$ yields the Hessian:
\begin{equation*}
    \mathrm{H}(\phi_i) = \left\{ \sum_{r=1}^{m_i} \left[\frac{-d_r}{h_{r,i}^2} - \frac{\eta_i - \Delta_i}{(1 - h_{r,i})^2}\right] \dot{h}_{r,i} \dot{h}_{r,i}^T
    + \left[\frac{d_r}{h_{r,i}} - \frac{\eta_i - \Delta_i}{1 - h_{r,i}}\right] \ddot{h}_{r,i} \right\}.
\end{equation*}
Finally the Fisher information matrix is:
\begin{equation*}
    \mathcal{I}(\phi_i) = -\mathrm{E}[\mathrm{H}(\phi_i)] = \sum_{r=1}^{m_i} \frac{\eta_r(r)}{h_{r,i}(1-h_{r,i})}\dot{h}_{r,i} \dot{h}_{r,i}^T.
\end{equation*}

Writing $w_r = \eta_i(r)/\{h_{r,i}(1-h_{r,i})\} \ge 0$, the Fisher information is $\mathcal{I}(\phi_i) = \sum_r w_r\,\dot{h}_{r,i}\dot{h}_{r,i}^{\top}$, a nonnegative combination of rank-one positive semi-definite matrices. Hence $v^{\top}\mathcal{I}(\phi_i)v = \sum_r w_r(\dot{h}_{r,i}^{\top}v)^2 \ge 0$, so $\mathcal{I}(\phi_i)$ is positive semi-definite. It is positive definite, and therefore invertible, if and only if the gradients $\{\dot{h}_{r,i} : w_r > 0\}$ span the parameter space; this requires at least $\dim\phi_i$ durations with $\eta_i(r) > 0$ and $0 < h_{r,i} < 1$. In that case $\Delta = \mathcal{I}(\phi_i)^{-1}U(\phi_i)$ is an ascent direction for $Q_{dwell,i}$, since $U^{\top}\Delta = U^{\top}\mathcal{I}^{-1}U > 0$ whenever $U \neq 0$. In practice a small ridge, $\mathcal{I} + \epsilon I$, guards against
near-singularity. The IRLS update for $\phi_i$ is then:
\begin{equation}
    \phi_i^{(k+1)} = \phi_i^{(k)} + \mathcal{I}(\phi_i^{(k)})^{-1}U(\phi_i^{(k)})
\end{equation}

The above equations entail all the necessary components for THSMM model fitting via the Baum-Welch algorithm. For details please refer to Appendix \ref{apx:hmmEmEquations}. For pseudo-code of the implementation please refer to \ref{apx:hmmEmPseudoCode}.

\begin{remark}[IRLS and EM guarantee]
\cite{DEMPSTER1977} show that the observed-data log-likelihood satisfies $\ell(\theta) - \ell(\theta^{(t)}) \geq Q(\theta \mid \theta^{(t)}) - Q(\theta^{(t)} \mid \theta^{(t)})$, so any update that does not decrease $Q$ also does not decrease $\ell$. This motivates the generalized EM (GEM) algorithm, which relaxes the M step from fully maximizing $Q$ to merely increasing it. IRLS, initialized at the current estimate $\theta^{(t)}$, satisfies this condition provided each step increases $Q$; this holds when $Q$ is concave in the regression parameters (e.g., under a canonical link) or can be enforced by step-halving. Hence IRLS need not be run to convergence, and even a single safeguarded step yields a valid GEM iteration.
\end{remark}

\section{Simulation Experiments}\label{sec:sim}

In the following sections we stress test the THSMM in a variety of settings. To focus on comparability and keep results concise we keep our attention to emission distributions of nonparametric $L^2$ mean curves with various sojourn distributions.
Further simulations can be found in the repository where we extensively test homogeneous mixtures of Brownian motion with linear drift, the Ornstein-Uhlenbeck process, and fractional Brownian motion with linear drift where the hidden process is a semi-Markov process. 
These simulations confirm that imposing the geometric distribution on 
the THSMM semi-Markov sojourn does computationally agree with the model
fitted by the THMM---via the \texttt{funHMM} CRAN package---as expected
mathematically.  They also demonstrate how the THSMM can outperform 
other models when the sojourn distribution is correctly specified 
or left as nonparametric.

For each scenario we perform $50$ replications. Each replicate has its own seed for data generation reproducibility. To evaluate model performance, we analyze overall performance accuracy and quality of parameter recovery. For performance accuracy, we use the adjusted Rand index (ARI) as our performance metric. The ARI is a popular method of measuring the agreement between two sets of labels and is computed in R via the \texttt{adjustedRandIndex()} function in the \texttt{mclust} package \citep{MCLUST}. An ARI value of 1 indicates a perfect match whereas an ARI value of 0 indicates random guessing. We secondly compare clustering accuracy using cross entropy as an alternative measurement. To asses the parameter recovery, we calculate bias, standard deviation, and root mean squared error (RMSE), of fitted values from the 50 replications. Note that this is not a bootstrapped estimate as a fresh synthetic data set is generated from the same set of base parameters for each replicate.

We point the reader to the repository for code to build and fit a THSMM as well as further simulations (\href{https://github.com/JakeHoover/thsmm}{repository link}). 

\subsubsection*{Data Generation Procedures}

\begin{figure}
    \centering
    \includegraphics[width=0.8\linewidth]{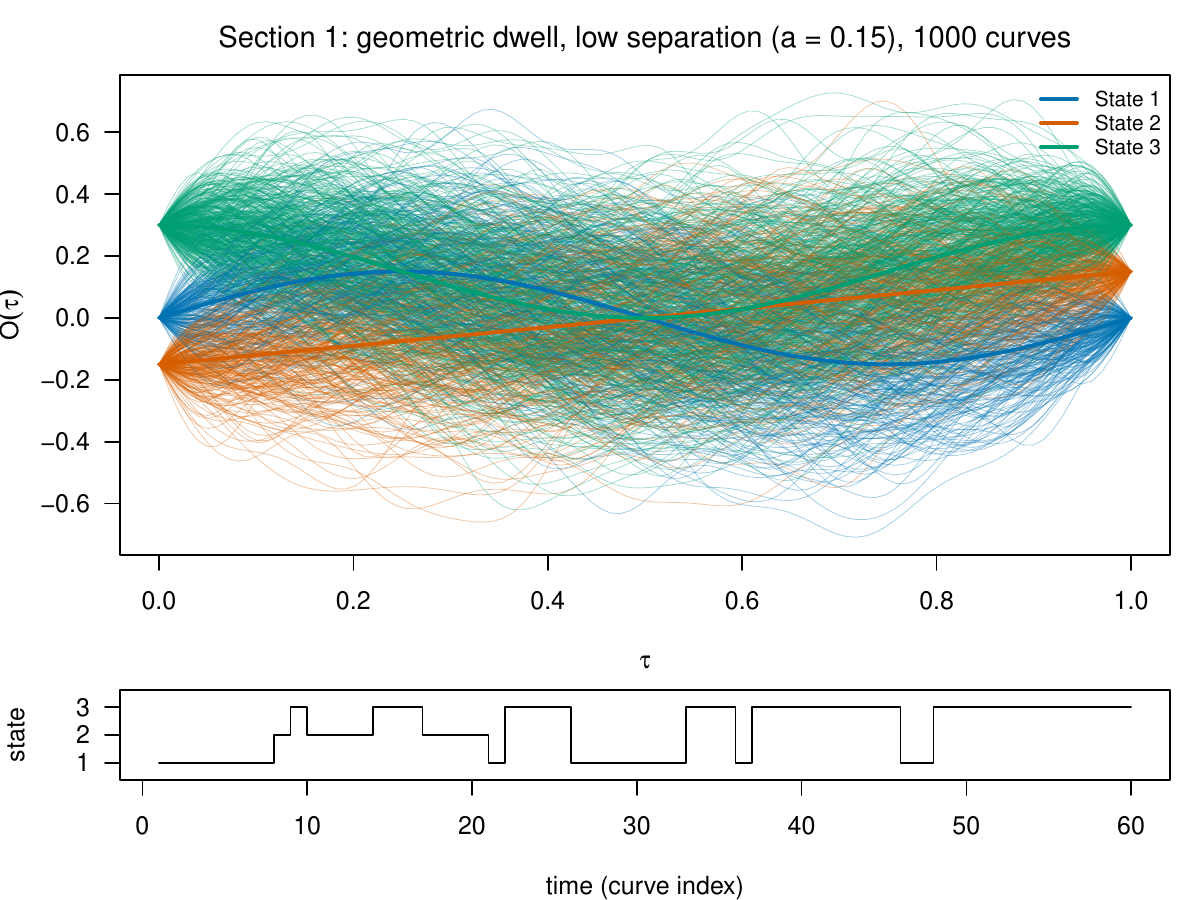}
    \caption{Synthetic non parametric $L^2$ curves generated with low separation.}
    \label{fig:synthDatGeoLow}
\end{figure}

 Each dataset has three states and transitions are uniform across states (jump from state $i$ to $j \neq i$ with probability $1/2$ for $i,j \in \{1,2,3\}$). Given its hidden state $j$, each observation is a curve on $100$ equispaced points of $[0,1]$ of the form
\begin{equation*}
  O(\tau) = h_j(\tau) + \sum_{m=1}^{16} \xi_m \, \frac{\sqrt{2}\,\sin(m\pi\tau)}{m\pi}, \qquad \xi_m \overset{\text{iid}}{\sim} N(0, 0.4^2),
\end{equation*}
so that the noise is a smooth Brownian bridge truncated to $16$ terms of its Karhunen--Lo\`eve expansion and is the same for every state. The state mean curves are $h_1(\tau) = a\sin(2\pi\tau)$, $h_2(\tau) = a(2\tau - 1)$ and $h_3(\tau) = a\{\cos(2\pi\tau) + 1\}$, where $a = 0.15$, $0.2$ and $0.3$ give low, medium and high separation between the states. Each of the proceeding sections then modifies the sojourn distributions for each state in order to better test the question at hand. 
For Sections \ref{sec:thmmRecovery} and \ref{sec:sojournRecovery} use a homogeneous mixture of sojourns and in Section \ref{sec:compSynthMods} we also consider a heterogeneous mixture of sojourns where each state has its own unique distribution. See Figure \ref{fig:synthDatGeoLow} for an example of a synthetic data set where we used geometric sojourn distributions with low separation between observations within states.

\begin{remark}[Mixtures of Sojourn Distributions]
The current implementation of THSMM supports the choice of a unique sojourn distribution for each state; 
e.g. the user could specify shifted Poisson, Geometric, and Negative Binomial for states 1, 2, and 3, respectively. 
In practice, this is useful as each family has its own benefits. For example, if in our data we see two states with long and consistent sojourn distributions then we may want to use a Shifted Poisson or a Negative Binomial for those states. But if we see that a specific state has a very unstable sojourn then it may be the case that a nonparametric or a geometric sojourn is more suitable. Thus the combination of sojourns will yield the most natural parameterization. This obviously comes at the expense of doing quite a bit of data analytics before fitting the model (but this is often the case anyways). Future work is aimed at building initialization procedures that can suggest or even automatically choose the the most suitable sojourn distribution for each state.
\end{remark}

\subsection{THMM Recovery}\label{sec:thmmRecovery}

\begin{figure}
    \centering
    \includegraphics[width=\linewidth]{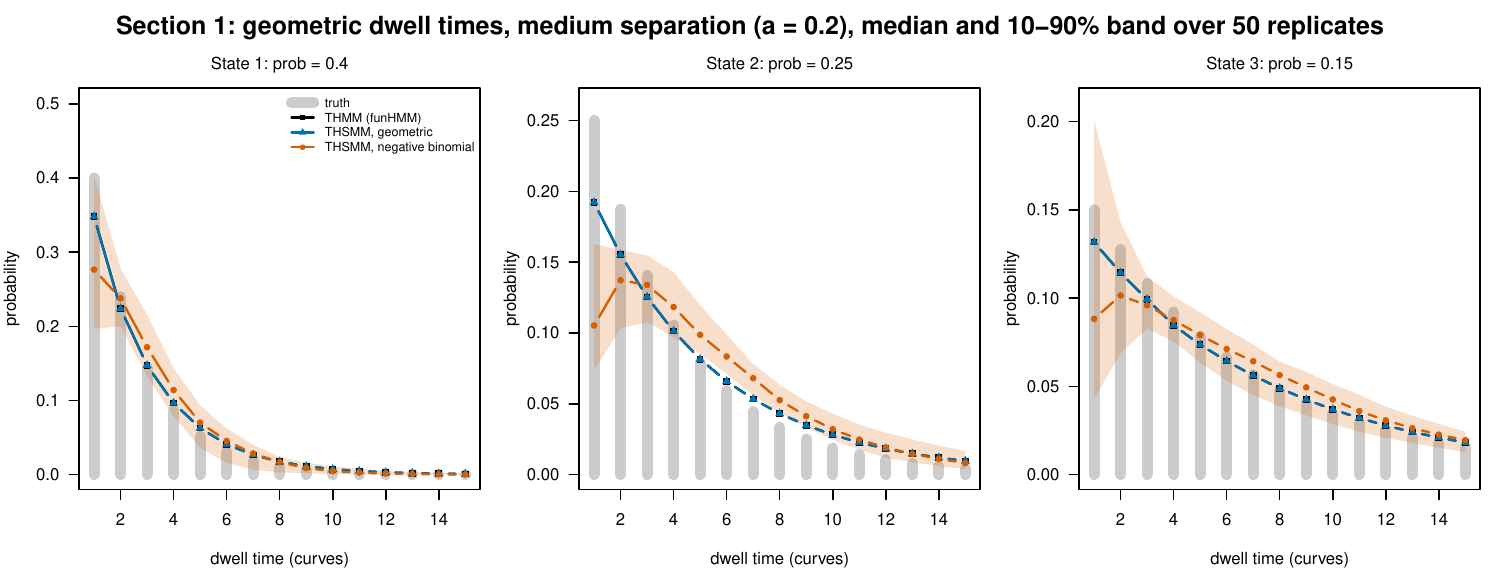}
    \caption{Fitted sojourn distribution per state for THMM, and THSMMs on nonparametric $L^2$ curves with geometric sojourn distributions.}
    \label{fig:geoRecovery}
\end{figure}

A natural question to ask is if the THSMM can recover a THMM. To this extent we generate synthetic data where each state has a geometric sojourn distribution. We then fit three models, a THMM, a THSMM with homogeneous geometric sojourn distributions, and a THSMM with homogeneous negative binomial distributions, where an aggregate size of $m=30$ was used for each state of the THSMM. In this sense the THSMM with geometric sojourns should be able to exactly recover the THMM and the THSMM with negative binomial should be able to get quite close, as a geometric distribution is a special case of a binomial distribution. See Table \ref{tab:geomTruth} for various statistics of the fitted models over 50 replications. 
As expected, the THSMM reproduces the THMM exactly when geometric sojourns are used, see Figure \ref{fig:geoRecovery} for plots of the fitted sojourns. 

An interesting takeaway is that separation of the observations drives the estimation quality. As seperation increases, the total variation (TV) distance between the fitted and the true dwell pmfs falls from about $0.22-0.25$ at low separation, to $0.10-0.12$ at medium, to under $0.04$ at high. Root means square error (RMSE) falls similairly, from $0.04-0.15$ to below $0.011$. Suggesting --- at least in the case of geometric sojourn distributions --- the when states overlap, a short visit to a state may get absorbed by a neighboring state, conversely long visits may unnecessarily transition out of a state; 
this is a known consequence for HMMs in a continuous and discrete data setting but now highlighted clearly for functional data. 

Fitting the negative binomial when the truth is known to be geometric highlights how the THSMM behaves with an unnecessarily flexible dwell family. At low separation the size parameter estimates rance from $15.6$ to $10^{11}$ with standard deviation estimates much larger. Meaning that in some replicates the size parameter diverges towards the Poisson limit. At medium separation the size estimates begin to stabilize but remain upwardly biased ($1.5-2.2$ against a true value of $1$) and at high separation they are close to the truth ($1.12-1.24)$. The extra flexibility does not noticeably effect the fitted emission distributions as the mean curve RMSEs are almost identical to the geometric fits. The cost is a modest loss of precision in the dwell distribution and an unidentified size parameter when separation is low. 

In conclusion, the THSMM recovers the THMM exactly when the dwell family is geometric. When the dwell family is more flexiable that it needs to be, the THSMM still estimates mean dwells and state mean curves as well as the correctly specified model.

\subsection{THSMM Sojourn Recovery}\label{sec:sojournRecovery}

We now seek to answer the main question at hand, if the THSMM can recover the true sojourn distribution when we know that it is not geometric. Here we simulate nonparametric $L^2$ curves with low, medium, and high separation using two sojourn schemes, a shifted Poisson%
with shift $(3,2,4)$ for states $1,2$, and $3$, respectively, and negative binomial. See Table \ref{tab:nongeomTruth} for various statistics of the fitted models over 50 replications. Note, we refer to the ``true model'' as the one whose sojourn distributions match that of the ones used to generate the data.  

\begin{figure}
    \centering
    \includegraphics[width=\linewidth]{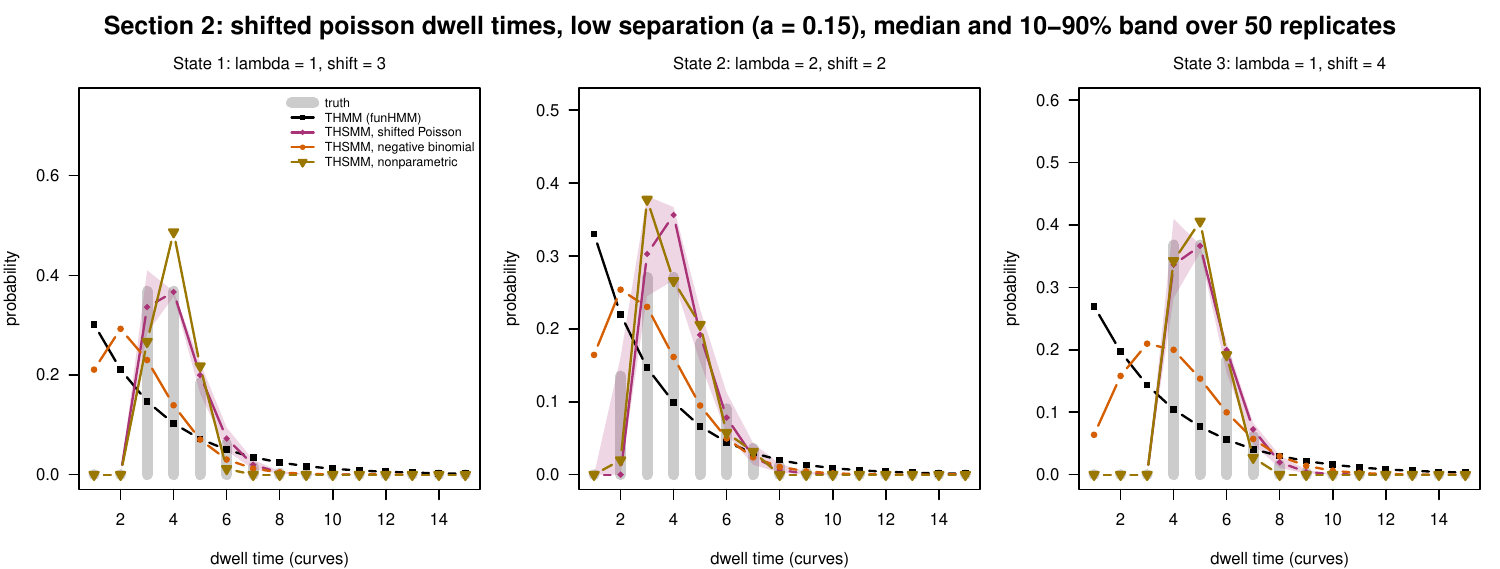}
    \caption{Fitted sojourn distributions per state for THMM and various THSMMs on non parametric $L^2$ curves with shifted Poisson sojourn distributions. In pink the estimated $10-90\%$ band over 50 replications of the fitted Poisson sojourns.}
    \label{fig:poisRecovery}
\end{figure}

The low separation setting with shifted Poisson dwell times shows most clearly how much the dwell specifications matter when the states are hard to tell apart, see Figure \ref{fig:poisRecovery}. All three true distributions put no mass below a minimum sojourn length of $s_j \in \{3,2,4\}$. As expected the THMM cannot represent this minimum and underestimates every mean dwell time giving $(3.25,2.96,3.68)$ against the true values of $(4,4,5)$. The negative binomial family also has poor performance with its mean dwell times having a downwards bias, due to the constraint of needing to put some mass in the first four dwell times. The non-parametric sojourn recovers mean dwell very accurately however at the cost of having a much higher TV distance, which is expected when estimating an unstructured pmf. Finally, the shifted poisson gives the most accurate fit, it recovers the shift for each state within $0.02$, and estimates every man dwell to within $0.09$. 

\begin{figure}
    \centering
    \includegraphics[width=\linewidth]{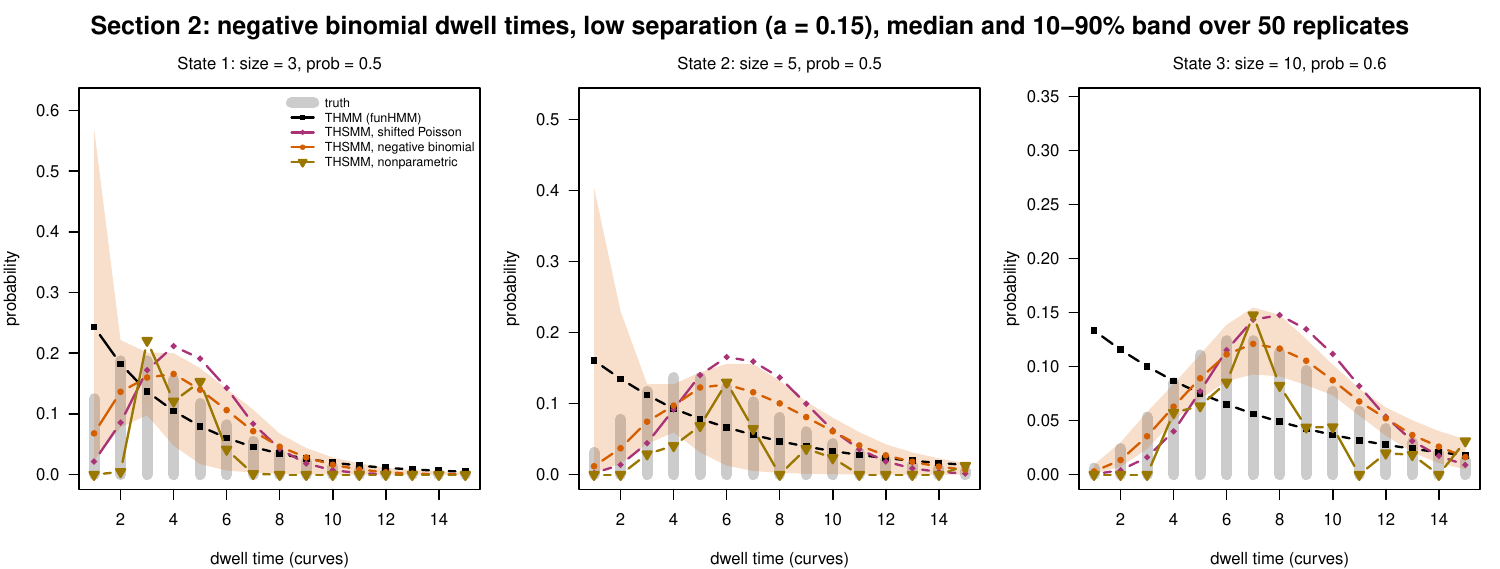}
    \caption{Fitted sojourn distributions per state for THMM and various THSMMs on non parametric $L^2$ curves with negative binomial sojourn distributions. In orange the estimated $10-90\%$ band over 50 replications of the fitted negative binomial sojourn.}
    \label{fig:nbRecovery}
\end{figure}

In the low separation setting when the true dwell times are negative binomial, there is no definitive winner between models. The true distributions are overdispersed and lie much closer to the geometric family than the shifted Poisson, leaving more ``wiggle room'' for various sojourn distributions to fit, see Figure \ref{fig:geoRecovery}. The shifted Poisson THSMM trades bias for stability as its shape is completely dictated by a single parameter. The cost comes at an upwards bias towards mean dwell times where they were overestimated by $0.68-0.93$. However, this added stability allows for it to produce the most accurate emission distributions from among the models. By assigning zero probability to sojourns shorter than $\hat s_j$, it suppresses spurious short state visits in the forward–backward recursions. The posterior state weights used in the emission update are therefore less contaminated by observations from overlapping states. Non-parametric sojourn distributions struggle in this setting as overlapping states plague the increased flexibility of the distribution with numerical instability. 
In Figure \ref{fig:nbRecovery}, the nonparametric sojourns are quite rough and have erratic behaviors. Finally, the negative binomial distribution recovers the shape of the sojourn the best. However, individual parameters are poorly identified. In some cases the size degenerated to the Poisson limit and the $p_j$ estimates are biased upwards to compensate. This can be seen in the $10-90\%$ bands in the plot, with the large spike indicates a misspecified fit. 

Many more results can be found in the accompanying repository.
First, the negative binomial cannot recover the under-dispersion that the shifted Poisson can tackle; 
conversely, the shifted Poisson is typically upwardly biased in mean state duration estimates. 
Second, the nonparametric sojourn distribution estimates become more stable as the separation of the data increases, see Table \ref{tab:nongeomTruth}. 
This suggests that more data could also remedy stability. 
Finally, the importance of running many repetitions to achieve a model with adequately stable estimates is necessary when data separation is low. 

In conclusion, The THMM’s geometric assumption creates a dwell error of known size that more data cannot remove. Any non-geometric THSMM reduces it, and a correctly specified family nearly eliminates it at moderate separation. The nonparametric THSMM is a useful default that makes no assumption about form, provided the states are reasonably well separated. When they are not, a parametric family with the right dispersion direction is the better choice. When the states are hard to distinguish, the dwell model also improves the recovered state mean curves.

\subsection{THSMM Comparative Performance}\label{sec:compSynthMods}

We investigate Viterbi preformance between THMM and THSMM models. This is of quite some importance as it can be argued that the actual usefulness of either of these models (or any latent state space model) is to decode the hidden state sequence of observations. As before we simulate nonparametric $L^2$ curves at low, medium, and high separation, under shifted Poisson and negative binomial sojourn distributions. Additionally to form a more comprehensive study we further simulate Brownian motion (BM), Ornstein-Uhlenbeck (OU) Process, and fractional Brownian Motion (fBM), under shifted Poisson, negative binomial, Weibull and a mixed family setting where each of the preceding is the sojourn for an individual state. 

See Table \ref{tab:ariL2Dwell} for ARIs of the nonparametric $L^2$ curve setting used in the previous subsections. First, the most meaningful gains to ARI occurred in the low separation setting. This is where the semi-Markov structure is most useful. When states overlap, a THSMM’s probability of leaving a state depends on how long the process has already spent there, so an ambiguous observation early in a sojourn is attributed to the current state rather than treated as evidence of a transition. The mechanism is clearest through the dwell hazard
\begin{align*}
    c_j(r) = P(D_j = r \mid D_j \ge r) = \frac{d_j(r)}{1 - \sum_{u<r} d_j(u)},
\end{align*}
which gives the probability of leaving state $j$ after $r$ time steps, given that it has not left before then. In the expanded-state representation, $c_j(r)$ is precisely the exit probability from the $r$-th substate of state $j$’s aggregate. 
At each time step, the forward–backward recursions weigh this prior exit probability against the emission evidence: the posterior log-odds of a transition combine $\log\{c_j(r)/(1 - c_j(r))\}$ with the log-likelihood ratio of the observation under the competing states. When states overlap, that likelihood ratio is close to zero, so the dwell term dominates the decision. 
As observations become more well separated the semi-Markov dynamics play less of a roll and emission densities take over the state assignment from the Viterbi algorithm. This is not a detriment to the THSMM. Classically best model fitting practice suggest to fit a model with the fewest number of parameters while not sacrificing accuracy. Therefore, in a setting where the data is well separated the user should take an Occam's razor approach to model fitting (which is historically the most commonly accepted approach) and choose a THMM instead of a THSMM.

\begin{table}
\centering
\footnotesize
\setlength{\tabcolsep}{5pt}
\caption{Viterbi ARI against the true states (mean (sd) over 50 replicates) by
dwell truth and separation. THSMM-G is the THMM fitted by the THSMM code
(geometric dwell); THSMM-true uses the true dwell family.}
\label{tab:ariL2Dwell}
\begin{tabular}{@{}llccc@{}}
\toprule
Dwell truth & Sep. & THMM (funHMM) & THSMM-G & THSMM-true \\
\midrule
Negative binomial & low    & 0.665 (0.131) & 0.665 (0.131) & 0.710 (0.137) \\
                  & medium & 0.950 (0.017) & 0.950 (0.017) & 0.948 (0.018) \\
                  & high   & 0.997 (0.003) & 0.997 (0.003) & 0.995 (0.003) \\
\midrule
Shifted Poisson   & low    & 0.519 (0.091) & 0.519 (0.091) & 0.864 (0.031) \\
                  & medium & 0.950 (0.015) & 0.950 (0.015) & 0.970 (0.010) \\
                  & high   & 0.998 (0.002) & 0.998 (0.002) & 0.999 (0.002) \\
\bottomrule
\end{tabular}
\end{table}

To extend the study and confirm that gains are not made solely on the emission emission distributions, we repeat the comparison with BM, OU, and fBM, emission distributions as well. See Figures \ref{fig:bmPois}, \ref{fig:bmMix}, \ref{fig:ouNb}, and \ref{fig:fbmW} for examples of plotted emission distributions with various sojourn distributions. 

The oracle choice for the sojourn distribution is optimal,
which is seen in all twelve process–scenario combinations
in Table~\ref{tab:ari-process}.
Nevertheless, 
the nonparametric THSMM keeps most of this advantage without knowing the dwell family. Its gains are $0.10$–$0.14$ for shifted Poisson, $0.03$–$0.05$ for Weibull and $0.02$–$0.06$ for mixed families, typically $0.01$–$0.03$ below the true family. The exception is the peaked negative binomial. There the nonparametric THSMM is no better than the THMM: its mean differences range from $-0.018$ to $0.007$, and it is better in only $34$–$52\%$ of replicates. This matches Section 2, where the nonparametric model struggled to estimate diffuse negative binomial dwell distributions. Overall, a correctly specified parametric family gives the best decoding. When the family is unknown, the nonparametric THSMM is a safe default that never decodes meaningfully worse than the THMM and is often substantially better.

\begin{table}
\centering
\footnotesize
\setlength{\tabcolsep}{4pt}
\caption{Viterbi ARI under non-geometric dwell times, BM / OU / fBM emissions at
low separation (mean (sd) over 50 replicates). THSMM-G is the THMM fitted by the
THSMM code (geometric dwell); THSMM-NP is nonparametric; THSMM-true uses the true
dwell family. Peaked NB = peaked negative binomial; Weibull has increasing hazard.
Failed fits are excluded.}
\label{tab:ari-process}
\begin{tabular}{@{}llcccc@{}}
\toprule
Process & Scenario & THMM & THSMM-G & THSMM-NP & THSMM-true \\
\midrule
BM  & Shifted Poisson & 0.640 (0.059) & 0.647 (0.055) & 0.760 (0.055) & 0.789 (0.048) \\
    & Peaked NB       & 0.681 (0.054) & 0.688 (0.058) & 0.669 (0.054) & 0.709 (0.064) \\
    & Weibull         & 0.678 (0.059) & 0.685 (0.058) & 0.718 (0.049) & 0.724 (0.050) \\
    & Mixed families  & 0.678 (0.047) & 0.677 (0.050) & 0.718 (0.052) & 0.739 (0.052) \\
\midrule
OU  & Shifted Poisson & 0.657 (0.046) & 0.661 (0.052) & 0.758 (0.051) & 0.786 (0.051) \\
    & Peaked NB       & 0.677 (0.066) & 0.687 (0.052) & 0.683 (0.054) & 0.723 (0.041) \\
    & Weibull         & 0.679 (0.043) & 0.681 (0.044) & 0.717 (0.048) & 0.723 (0.048) \\
    & Mixed families  & 0.655 (0.078) & 0.654 (0.092) & 0.711 (0.063) & 0.729 (0.073) \\
\midrule
fBM & Shifted Poisson & 0.631 (0.064) & 0.632 (0.059) & 0.767 (0.051) & 0.793 (0.047) \\
    & Peaked NB       & 0.674 (0.052) & 0.673 (0.058) & 0.676 (0.057) & 0.703 (0.068) \\
    & Weibull         & 0.671 (0.059) & 0.677 (0.057) & 0.721 (0.044) & 0.727 (0.047) \\
    & Mixed families  & 0.672 (0.066) & 0.676 (0.065) & 0.698 (0.072) & 0.748 (0.060) \\
\bottomrule
\end{tabular}
\end{table}

In conclusion, the advantage of modeling the sojourn distribution comes from the dwell structure, not from any particular emission process. And we can confirm that the THSMM even if misspecified will give no worse decoding accuracy than the THMM when the sojourn distributions are not geometric. 

\subsection{Summary}\label{sec:synthSummary}

Taken together, the simulations support three conclusions. First, the THSMM is a strict generalisation of the THMM. With geometric dwell times, it reproduces the THMM’s parameter estimates and Viterbi paths exactly, and fitting a more flexible family than needed costs only a modest loss of precision in the dwell distribution, with no effect on the mean dwells or state mean curves. Second, the THMM’s geometric assumption causes a dwell error that more data cannot remove. Any non-geometric THSMM reduces this error, and a correctly specified family nearly eliminates it at moderate separation. Misspecified parametric families still improve on the THMM, especially when their dispersion is in the right direction. The nonparametric THSMM is a robust default when the family is unknown, but it needs reasonable separation to estimate diffuse distributions. Third, these gains carry over to state decoding. The size of the gain depends on how far the dwell distribution is from geometric and on how hard the states are to distinguish. Under the shifted Poisson truth at low separation, the mean ARI rises from $0.52$ to $0.86$. Under Brownian motion, Ornstein–Uhlenbeck and fractional Brownian motion emissions, the true-family THSMM decodes significantly better than the THMM in every scenario. The common mechanism is the dwell hazard. When emissions overlap, a dwell distribution with an increasing hazard suppresses the spurious short visits that the geometric model favours a priori. This improves both segmentation and emission estimation, and the advantage fades only once the emissions alone separate the states.

\section{Pediatric Obstructive Sleep Apnea Data}\label{sec:osaData}

The gold standard for diagnosis of obstructive sleep apnea in children is by overnight polysomnography in a hospital or sleep clinic. Polysomnography provides multi-channel time series including an electroencephalogram (EEG), electrocardiography (ECG), electrooculography (EOG), and electromyography (EMG). Even for a single patient, this data is vast and should eventually be considered jointly to label sleep states and identify sleep disorders. For simplicity we choose one channel of EEG from one patient labeled as CF076 to be used as a proof of concept. This patient was selected from a group of seventy four pediatric patients with potential obstructive sleep apnea in a clinical study, Pro00057638, approved by University of Alberta. An in-depth analysis of this data can be found in \cite{CHALIFOUR2020}. The sampling rate for EEG is 512 samples per second. Each signal was split into a sequence of epochs, i.e. 30 second intervals. These signals were transformed into power spectral densities (PSD) using Welch’s method for each epoch. 
The reason for dividing time series into 30 second intervals is to group them with respect to five sleep stages: wake, rapid eye-movement (REM), and non-rapid eye-movement (NREM). 
The NREM category is further divided into three states: NREM1, NREM2, NREM3. The sleep stages are labelled per epoch by a sleep technician. For every sleep stage not labelled as wake, the patient is considered to be asleep. The spectral densities for each epoch may then be used to identify possible underlying states such as the sleep stages. There were 948 epochs in patient CF050.

Markov models and HMMs have a long history of being applied to the modelling of sleep states \citep{Zung1966, Yang1973, Kemp1986}.
The work of \cite{Penny1998} considered an HMM with Gaussian observations
for EEG analysis. 
This was followed by similar analyses in \cite{Flexer2002, Flexer2005}. More recent work includes \cite{Doroshenkov2007, Pan2012, Chen2015}. 
Typically, these approaches are based on various discretization and feature selection methods whose output is then fed into the classic HMM. The novelty of the THMM \citep{KASHLAK_THMM} is that it takes the entire power spectral density curve into account and hence obviates the need for such feature selection steps. 
However it still restricts modelling to the geometric sojourn distribution whereas the recent work of \cite{KASHLAK_FISCHER2026} suggest that the negative binomial sojourn may be more appropriate for modeling sleep cycles. 

We begin by fitting various THSMMs and THMMs, focousing on THSMMs with Shifted Poisson sojourn distributions and nonparametric functional emission distributions. 
We choose Shifted Poisson for two reason; first, both models have the same number of free parameters thus they only vary in the shape of the sojourn distribution; second, we found that throughout the exploratory process the Shifted Poisson has the most stable fitted parameters likely due to the fact that the distribution is parameterized only by the mean.

To vary how strongly the emissions drive the posterior relative to the dwell model, we scale the emission variance by a multiplier $c \in \{1, 0.3, 0.1 \}$, taking $\sigma_c = \sigma_1/\sqrt{c}$. 
Smaller $c$ flattens the emission likelihood allowing for a larger contribution of the sojourn distributions\footnote{Recall, the emission is modeled as a shift of a common centered Gaussian measure $O_t = h_k + \sigma Z_t$. The state mean $h_k$ is estimated nonparametrically over the Cameron-Martin norm (taken to be the second derivative). The emission is then the Onsager-Machlup functional $\log b_k(O_t) \approx -\frac{1}{2\sigma^2},\frac{1}{\Delta^3}\sum_{j}\big(\Delta^2 (O_t - h_k)\big)_j^2$, where $\Delta$ is the second difference over the 50 frequency bins. The EM does not fit $\sigma$. On a global scale it sets how sharply the emission term separates the states. Since $\log b_k$ is proportional to $1/\sigma^2$, this gives $\log b_k^{(c)}(O_t) = c\,\log b_k^{(1)}(O_t), b_k^{(c)} = \big(b_k^{(1)}\big)^{c}.$ For any two states $\frac{b_k^{(c)}(O_t)}{b_l^{(c)}(O_t)} = \left(\frac{b_k^{(1)}(O_t)}{b_l^{(1)}(O_t)}\right)^{c}$. As $c$ decreases, these ratios shrink to 1, the the emission term becomes flatter across states.}. 
Each model was fitted from 51 starts: a k-means baseline plus 50 random perturbations of it. We kept the fit with the highest likelihood and obtained a decoded state sequence with the Viterbi algorithm. We then matched the fitted states to the scored stages and compute accuracy and the adjusted Randi Index (ARI).

Both models recover sleep structure well beyond chance. Permuting the scored labels gives $\text{ARI} \approx 0.000$ (99th percentile 0.015). 
Every selected fit reaches $\text{ARI} \ge 0.32$ (moderate to weak) (see Table \ref{tab:eegVit}).

\begin{table}[ht]
    \centering{
    \caption{Viterbi agreement with the scored states for the highest likelihood fit at each $c$.}
    \label{tab:eegVit}
    \begin{tabular}{clccccc}
        \toprule
        $c$ & Model & Accuracy & Bal. acc. & ARI & Sojourns (scored: 71) & Epochs differing \\
        \midrule
        1.0 & THMM  & 0.532 & 0.487 & 0.325 & 101 & --  \\
        1.0 & THSMM & 0.528 & 0.478 & 0.323 & 95  & 8\% \\
        \midrule
        0.3 & THMM  & 0.545 & 0.503 & 0.331 & 49  & --  \\
        0.3 & THSMM & 0.614 & 0.510 & 0.431 & 66  & 60\% \\
        \midrule
        0.1 & THMM  & 0.548 & 0.490 & 0.397 & 24  & --  \\
        0.1 & THSMM & 0.609 & 0.504 & 0.441 & 55  & 52\% \\
        \bottomrule
    \end{tabular}}
\end{table}

\begin{figure}
    \centering
    \includegraphics[width=\linewidth]{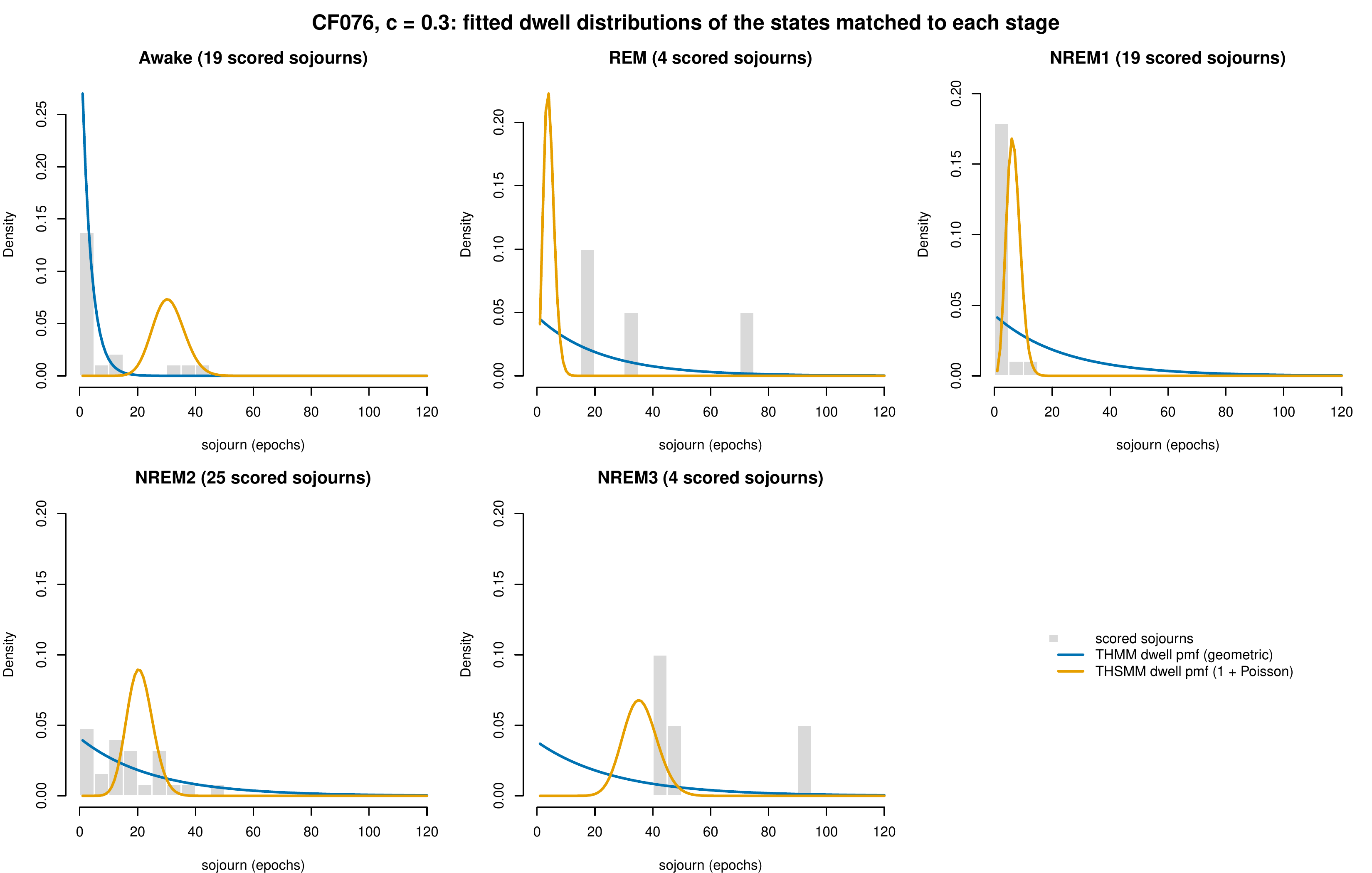}
    \caption{Fitted emission distributions for the THSMM when $c=0.3$.}
    \label{fig:c03fittedDwell}
\end{figure}

We see that the dwell time modeling matters most when emission are not decisive. 
At $c=1$, the emission dominates. The two Viterbi paths differ on only $8\%$ of epochs, and the THSMM and THMM are indistinguishable. As $c$ decreases, the dwell model shapes the decoded path more. 
The THSMM then improves accuracy by 6–7 points and ARI by 0.04–0.10. The THMM responds to weaker emissions by collapsing into very few long runs and its fitted mean dwell for NREM1 grows from 15 to 48 epochs against a scored 2.7. Notice the 49 sojourns at $c=0.3$ and 24 at $c=0.1$ (see Table \ref{tab:eegVit}).  
The THSMM keeps a sojourn count close to the scored one (66 at $c=0.3$). It also fits short NREM1 sojourns (6.7 epochs) together with long NREM2 and NREM3 sojourns (20.9 and 35.7). 
A single geometric rate per state cannot produce that contrast. 
See Figure \ref{fig:c03fittedDwell} for plots of the fitted sojourn distributions, were we see that the Poisson sojourn is adapting much better to dwells of longer lengths than that of the geometric. 
See \ref{apx:osa} for plotted Viterbi state sequences for the fitted models when $c=1.0,0.3,0.1$. 

\begin{figure}
    \centering
    \includegraphics[width=\linewidth]{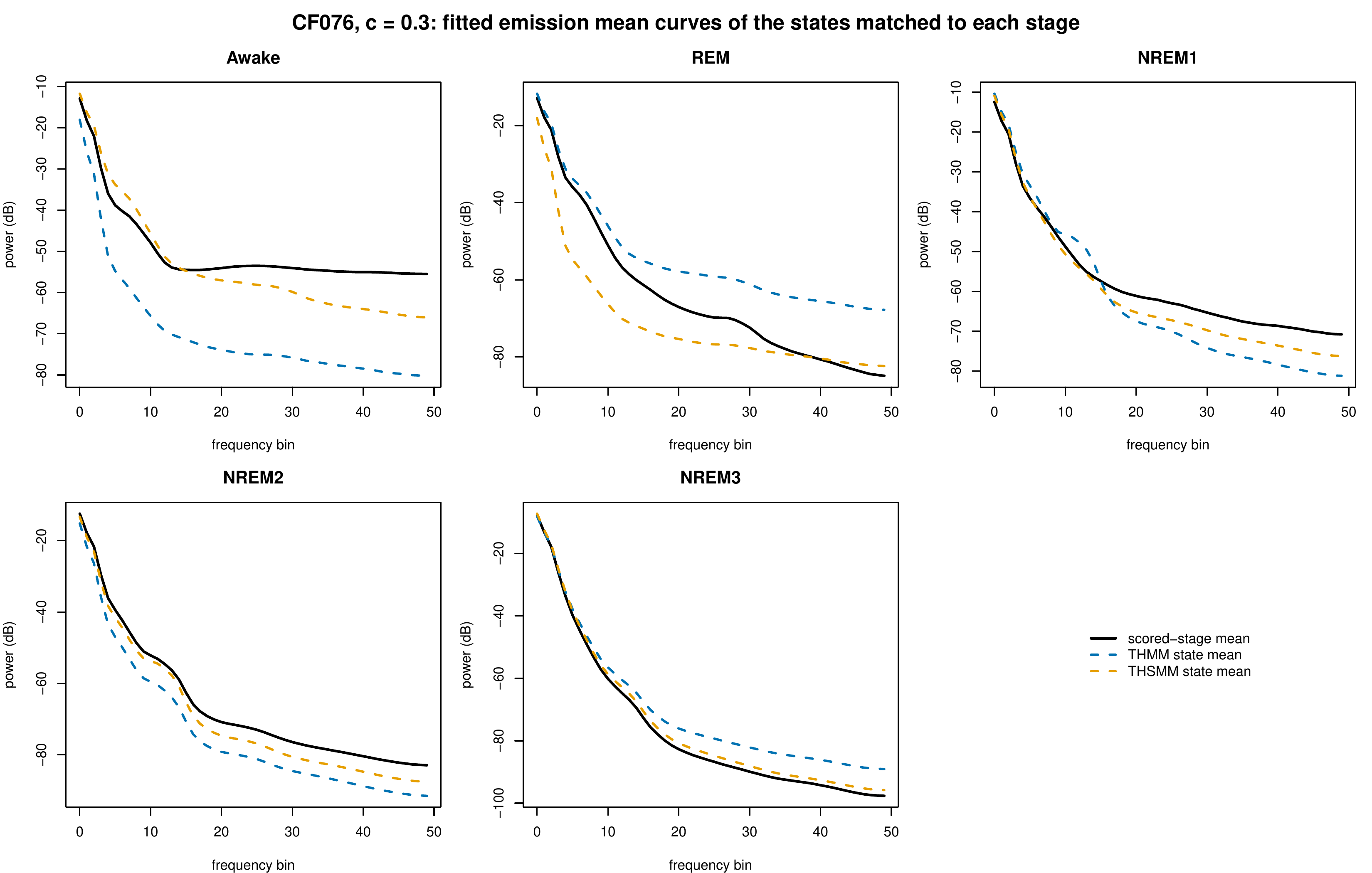}
    \caption{Fitted emission distributions for the THMM and THSMM when $c=0.3$.}
    \label{fig:c03fittedEmissions}
\end{figure}

\begin{figure}
    \centering
    \includegraphics[width=\linewidth]{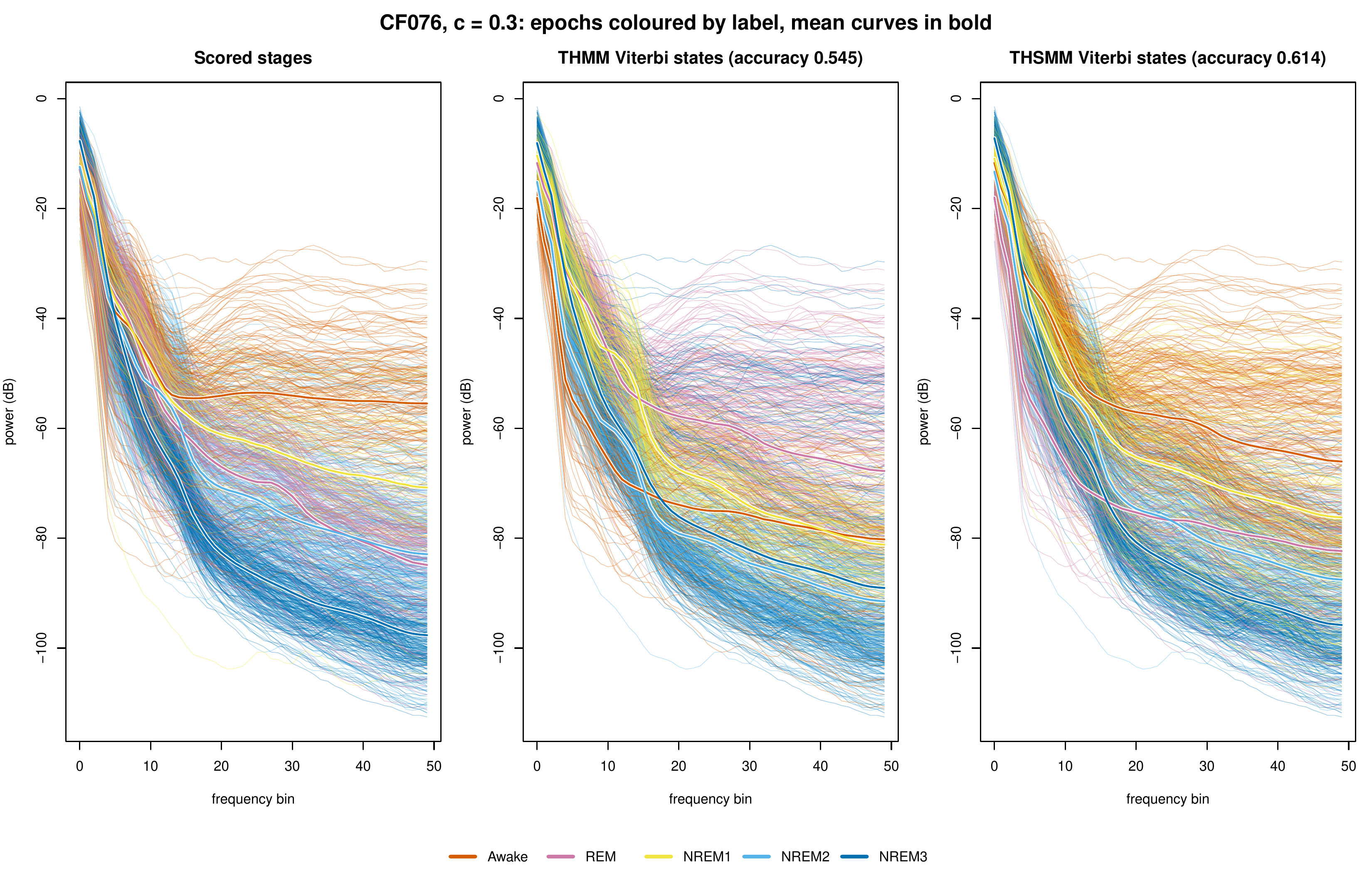}
    \caption{Observations labelled by score and predicted state from Viterbi for the THMM and THSMM when $c=0.3$.}
    \label{fig:c03datVit}
\end{figure}

Both models identify NREM3 well (recall 0.73–0.96). Neither recovers NREM1, which is short, rare ($5\%$ of epochs) and has a PSD closely resembling the states around it (recall $\le 0.24$). 
Most of the THSMM's improvement comes from NREM2, whose recall rises from about 0.41 to 0.68–0.72 at $c \le 0.3$. 
The main persistent error is between Awake and REM, whose smoothed PSDs are similar and is the major cause of low ARI scores for both models. Whichever of the two the models label correctly, most epochs of the other are assigned to it: REM is mostly labelled Awake by the THSMM at $c=0.3$, and Awake is mostly labelled REM in most of the other fits. 
A single EEG channel is unlikely to separate these stages without EOG or EMG. 
For plots of the fitted emission distributions see Figure \ref{fig:c03fittedEmissions} and see Figure \ref{fig:c03datVit} for fitted emissions overlaid on the data (where observations are coloured by decoded states from the specified model). 

In summary, replacing the sojourn dynamic gives no benefit when the functional emissions alone determine the path. When we temper the emission distributions, the sojourn distributions have a meaningful impact on likelihood and yield a better calibrated model with higher agreement with the technician scoring. The moderate ARI scores can likely be improved by implementing multiple signals into the emission distributions and is something to consider in future work.

\section{Conclusion}\label{sec:conclusion}

We develop a new model framework THSMM, by incorporating a semi-Markov component to the already existing THMM. By embedding the semi-Markov chain into the THMM transition probability matrix, we obtain Markov chain dynamics whose state durations can follow arbitrary distributions, approximating any known duration family up to length $m$. Furthermore, the implementation of THSMM is greatly simplified due to this framework as we have immediate access to the standard (T)HMM toolkit such as the Viterbi algorithm; this can be extended to further results such as THSMMs with covariates, in the state dependent process or the latent process, something we look forward to implementing in the future. Our EM framework yields closed-form updates for the functional distribution parameters during model fitting, a problem that is ill-posed for direct numerical optimization.

As demonstrated in the synthetic data setting the added flexibility yields a tangible improvement to decoded state accuracy; in particular when emission distributions have a strong overlap, the semi-Markov dynamics produce more stable Viterbi sequences. When applied to EEG data from patient CF076, a particularly challenging case to model, the THSMM again improves stability, achieving higher accuracy and ARI than its THMM counterpart.

The THSMM thus extends the THMM to arbitrary duration distributions at little additional computational cost, with clear gains in settings where state emissions overlap, and dwell times are not geometric.

\section*{Funding, Conflicts of interests, and AI Usage}

The authors would like to thank the Natural Sciences and Engineering 
Research Council of Canada (NSERC) and 
the McIntyre Memorial fund from the School of
Dentistry at the University of Alberta
for their grant funding in support of this research.
We declare that their are no conflicts of interest to report
in regard to this research project.  The ideas and work contained
in this paper came solely from the authors.  The original research
code was written by the authors in R and subsequently translated 
into C and optimized by Claude Opus.  This LLM written C code was 
carefully checked for accuracy and compared to the original R code 
by the authors.

\bibliography{refs,
              latexfiles/kasharticle,
              latexfiles/kashbook,
              latexfiles/kashmisc,
              latexfiles/kashpack,
              latexfiles/kashself}%

\newpage 

\begin{appendices}

\section{Derivations for EM Equations}\label{apx:hmmEmEquations}

In the following we note that the time $t = T$ is excluded. Although there has been much work dedicated to right censoring of semi-Markov processes (\cite{BullaThesis}). For ease of notation we denote $L = \mathbb{P}(x_{1:T})$ as the likelihood.

Derivation of $\eta_i(r)$: We want to find 
\begin{align*}
    \eta_i(r) = \sum_{t=1}^T \mathds{1} \{S_t = (i,r)\},
\end{align*}
but we do not have access to $S$ so we use the forward and backwards probabilities. 
\begin{align*}
    \mathrm{E}[\mathds{1}[S_t = (i,r)]|X_{1:T}] &= P(S_t = (i,r)|X_{1:T} = x_{1:T}) \\
    &= \frac{P(S_t = (i,r),x_{1:T})}{P(x_{1:T})} \\
    &= \frac{\alpha_t(i,r)\beta_t(i,r)}{L}
\end{align*}
Summing over all time then gives
\begin{align*}
    \eta_i(r) = \sum_{t=1}^{T-1} \frac{\alpha_t(i,r)\beta_t(i,r)}{L}
\end{align*}

Derivation of $\Delta_i(r)$: Begin with:
\begin{align*}
    P(X_{1:T},S_t=(i,r),S_{t+1}=(j,1)) = \alpha_t(i,r) \omega_{i,j} c_i(r) b_{t+1}(j,1) \beta_{t+1}(j,1)
\end{align*}
We do not care which $j$ we jump to only that we left state $i$, therefore we sum over $j\neq i$, finally summing over time and dividing by $L$ yields the expectation: 
\begin{align*}
    \Delta_i(r) &= \sum_{t=1}^T \frac{\alpha_t(i,r)c_i(r) \sum_{j\neq i} \omega_{i,j} b_{t+1}(j,1) \beta_{t+1}(j,1)}{L} \\
    &= c_i(r) \sum_{t=1}^T \frac{\alpha_t(i,r)G_t(i)}{L}.
\end{align*}

Derivation of $N(i,j)$: It is easiest to immediately start by taking the expected value of $N(i,j)$:
\begin{align*}
    \mathrm{E}&\left[\sum_{t=1}^T \sum_{r=1}^{m_i} \mathds{1}[S_t = (i,r), S_{t+1} = (j,1)]|X_{1:T} = x_{1:T}] \right] \\
    &= \sum_{t=1}^T \sum_{r=1}^{m_i} P(S_t = (i,r), S_{t+1} = (j,1)]|X_{1:T} = x_{1:T}) \\
    &=\sum_{t=1}^T \sum_{r=1}^{m_i} \frac{P(S_t = (i,r), S_{t+1} = (j,1), x_{1:T}) }{ L} \\
    &=\sum_{t=1}^T \sum_{r=1}^{m_i} \frac{\alpha_t(i,r) \omega_{i,j} c_i(r) b_{t+1}(j,1)\beta_{t+1}(j,1)}{L} \\
    &= \sum_{t=1}^T \frac{\omega_{i,j} b_{t+1}(j,1)\beta_{t+1}(j,1)F_t(i)}{L}
\end{align*}

Derivation of $\gamma_t(i,r)$: Corresponds to the regular EM update for a HMM except not we only focus on non zero entries using forward and backwards probabilities. Yielding: 
\begin{align*}
    \gamma_t(i,r)  = \frac{\alpha_t(i,r)\beta_t(i,r)}{L}
\end{align*}

Throughout all the derivations above the forward and backwards passes produce $\alpha$, $\beta$, $F$, and $G$, which are used to create $\eta_i(r)$, $d_i(r)$, and $N(i,j)$.

\section{EM Pseudo Code}\label{apx:hmmEmPseudoCode}

\providecommand{\nStates}{K}
\providecommand{\aggSize}[1]{m_{#1}}
\providecommand{\rplus}{r^{+}}

\begin{algorithm}[H]
\caption{The THSMM EM algorithm}\label{alg:thsmmEM}
\small
\begin{algorithmic}[1]
\Require Observations $x_1,\ldots,x_T$; number of states $\nStates$; for each
  state $i$ a dwell-time family $p_i(\,\cdot\,;\phi)$, aggregate size
  $\aggSize{i}$ and emission functional $f(\,\cdot\,;\theta)$; tolerance
  $\epsilon$.
\State Initialize $\delta_i$, \ $\omega_{i,j}$ (with $\omega_{i,i}=0$), \
  $\phi_i$, \ $\theta_i$ \ for $i,j=1,\ldots,\nStates$.
\Statex \textit{Substates are pairs $(i,r)$ with $r=1,\ldots,\aggSize{i}$, and
  $\rplus=\min(r+1,\aggSize{i})$.}
\Repeat \Comment{iteration $k=1,2,\ldots$}
  \Statex \textbf{Model quantities}
  \State $c_i(r) \gets p_i(r;\phi_i)\big/\{1-F_i(r-1;\phi_i)\}$
    for $r=1,\ldots,\aggSize{i}$
    \Comment{hazards, Eq.~\eqref{eq:hazard}}
  \State $\ell_t(i) \gets \log f(x_t;\theta_i)$, \
    $M_t \gets \max_i \ell_t(i)$, \
    $\tilde b_t(i) \gets \exp\{\ell_t(i)-M_t\}$
    \Comment{offset emissions}
  \Statex \textbf{Forward pass}
  \State $\tilde\alpha_1(i,r) \gets \delta_i\,\tilde b_1(i)\,\mathbf{1}\{r=1\}$;
    \ $u_1 \gets \sum_{i,r}\tilde\alpha_1(i,r)$; \
    $\hat\alpha_1 \gets \tilde\alpha_1/u_1$
  \For{$t=2,\ldots,T$}
    \State $\hat F_{t-1}(i) \gets \sum_{r=1}^{\aggSize{i}}
      \hat\alpha_{t-1}(i,r)\,c_i(r)$
      \Comment{exit flux, Eq.~\eqref{eq:F}}
    \State $\tilde\alpha_t(j,1) \gets \sum_{i\neq j}\omega_{i,j}\,\hat F_{t-1}(i)$,
      \ $\tilde\alpha_t(j,r) \gets 0$ for $r>1$
      \Comment{enter $j$}
    \State $\tilde\alpha_t(j,\rplus) \gets \tilde\alpha_t(j,\rplus)
      + \hat\alpha_{t-1}(j,r)\{1-c_j(r)\}$ for all $(j,r)$
      \Comment{remain in $j$}
    \State $\tilde\alpha_t(j,r) \gets \tilde b_t(j)\,\tilde\alpha_t(j,r)$; \
      $u_t \gets \sum_{j,r}\tilde\alpha_t(j,r)$; \
      $\hat\alpha_t \gets \tilde\alpha_t/u_t$
  \EndFor
  \State $\log L^{(k)} \gets \sum_{t=1}^{T}\left(\log u_t + M_t\right)$
    \Comment{$=\log\mathbb{P}(x_{1:T})$}
  \Statex \textbf{Backward pass} \quad $\hat\beta_T(i,r) \gets 1$ for all $(i,r)$
  \For{$t=T-1,\ldots,1$}
    \State $\hat e_t(j) \gets \tilde b_{t+1}(j)\,\hat\beta_{t+1}(j,1)\big/u_{t+1}$
      \Comment{value of entering $j$}
    \State $\hat G_t(i) \gets \sum_{j\neq i}\omega_{i,j}\,\hat e_t(j)$
      \Comment{exit value, Eq.~\eqref{eq:G}}
    \State $\hat\beta_t(i,r) \gets \{1-c_i(r)\}\,\tilde b_{t+1}(i)\,
      \hat\beta_{t+1}(i,\rplus)\big/u_{t+1} + c_i(r)\,\hat G_t(i)$
  \EndFor
  \Statex \textbf{Expected sufficient statistics}
  \State $\iota_t(i,r) \gets \hat\alpha_t(i,r)\,\hat\beta_t(i,r)$, \quad
    $\iota_t(i) \gets \sum_{r=1}^{\aggSize{i}}\iota_t(i,r)$
    \Comment{posterior occupancy}
  \State $\eta_i(r) \gets \sum_{t=1}^{T-1}\iota_t(i,r)$
    \Comment{expected visits}
  \State $\Delta_i(r) \gets c_i(r)\sum_{t=1}^{T-1}\hat\alpha_t(i,r)\,\hat G_t(i)$
    \Comment{expected departures}
  \State $N(i,j) \gets \omega_{i,j}\sum_{t=1}^{T-1}\hat F_t(i)\,\hat e_t(j)$
    \Comment{expected $i\to j$ jumps}
  \Statex \textbf{Maximization}
  \State $\delta_i \gets \iota_1(i,1)$; \
    $\omega_{i,j} \gets N(i,j)\big/\sum_{l\neq i}N(i,l)$ for $i\neq j$
    \Comment{$\omega$ fixed when $\nStates=2$}
  \For{$i=1,\ldots,\nStates$}
    \If{$p_i$ is nonparametric}
      \State $c_i(r) \gets \Delta_i(r)/\eta_i(r)$, \ $r=1,\ldots,\aggSize{i}$
        \Comment{no tail: $c_i(\aggSize{i})= 1$}
      \State $p_i(r) \gets c_i(r)\prod_{s<r}\{1-c_i(s)\}$
    \ElsIf{$p_i$ is geometric}
      \State $c_i \gets \sum_r \Delta_i(r)\big/\sum_r \eta_i(r)$
    \Else
      \State $\phi_i \gets \Call{DwellScoring}{\phi_i,\,\Delta_i,\,\eta_i}$
        \Comment{Algorithm~\ref{alg:dwellScoring}}
    \EndIf
    \State $\theta_i \gets \arg\max_{\theta}\sum_{t=1}^{T}\iota_t(i)\,
      \log f(x_t;\theta)$
      \Comment{Table~\ref{tab:emissionUpdates}}
  \EndFor
\Until{$k>1$ and $\left|\log L^{(k)}-\log L^{(k-1)}\right|
  < \epsilon\left(1+\left|\log L^{(k)}\right|\right)$}
\State \Return $\delta$, $\omega$, $\phi$, $\theta$
\end{algorithmic}
\end{algorithm}

\begin{algorithm}[H]
\caption{Dwell-time M-step for a parametric family (Fisher scoring)}
\label{alg:dwellScoring}
\begin{algorithmic}[1]
\Function{DwellScoring}{$\phi_i,\,\Delta_i,\,\eta_i$}
  \State $\psi \gets$ unconstrained transform of $\phi_i$
    \Comment{e.g.\ $\log\lambda$; shifts held fixed}
  \State Let $Q_{dwell,i}(\psi) = \sum_{r=1}^{\aggSize{i}}
    \Big[\Delta_i(r)\log h(r;\psi) + \{\eta_i(r)-\Delta_i(r)\}\log\{1-h(r;\psi)\}\Big]$
  \For{$s=1,\ldots,S$}
    \Comment{a few steps per EM iteration suffice}
    \State $h_{r,i} \gets h(r;\psi)$, \
      $\dot h_{r,i} \gets \partial h(r;\psi)/\partial\psi$, \
      for $r\in\mathcal{R}=\{r: \eta_i(r)>0,\ 0<h_{r,i}<1\}$
    \State $U \gets \sum_{r\in\mathcal{R}}
      \dfrac{\Delta_i(r)-\eta_i(r)\,h_{r,i}}{h_{r,i}(1-h_{r,i})}\,\dot h_{r,i}$
      \Comment{score}
    \State $\mathcal{I} \gets \sum_{r\in\mathcal{R}}
      \dfrac{\eta_i(r)}{h_{r,i}(1-h_{r,i})}\,\dot h_{r,i}\,\dot h_{r,i}^{\top}$
      \Comment{expected information}
    \State $\Delta\psi \gets \mathcal{I}^{-1}U$; \ $\lambda \gets 1$
    \While{$Q_{dwell,i}(\psi+\lambda\Delta\psi) < Q_{dwell,i}(\psi)$}
      \State $\lambda \gets \lambda/2$
        \Comment{step-halving keeps EM monotone}
    \EndWhile
    \State $\psi \gets \psi+\lambda\Delta\psi$
  \EndFor
  \State \Return $\phi_i(\psi)$
\EndFunction
\end{algorithmic}
\end{algorithm}

\FloatBarrier
\begin{table}[!htbp]
\centering
\caption{Weighted M-step updates
  $\theta_i=\arg\max_\theta\sum_{t=1}^{T}\iota_t(i)\log f(x_t;\theta)$ in
  Algorithm~\ref{alg:thsmmEM}, where $W_i=\sum_{t=1}^{T}\iota_t(i)$. Each
  entry gives $\log f(x_t;\theta_i)$ up to terms free of $\theta_i$, then the
  update.}
\label{tab:emissionUpdates}
\small
\renewcommand{\arraystretch}{1.3}
\begin{tabular}{@{}l l@{}}
\toprule
Emission & Functional and update \\
\midrule
\begin{tabular}[t]{@{}l@{}}Brownian motion\\ with drift; fractional\\ BM (Hurst $\nu$)\end{tabular}
  & $\begin{aligned}[t]
      \log f(x_t;\theta_i) &= \theta_i S_t-\tfrac12\theta_i^2 Q\\
      \hat\theta_i &= \frac{\sum_t\iota_t(i)\,S_t}{Q\,W_i}
    \end{aligned}$ \\ \noalign{\medskip}
\begin{tabular}[t]{@{}l@{}}Ornstein--Uhlenbeck,\\ mean $b_{0i}/b_{1i}$\end{tabular}
  & $\begin{aligned}[t]
      \log f(x_t;\theta_i) &= b_{0i}Z_t-b_{1i}I_t\\
        &\quad-\tfrac12\big(b_{0i}^2 s_L-2b_{0i}b_{1i}M_t^{(1)}+b_{1i}^2M_t^{(2)}\big)\\
      \hat b_{1i} &= \max\left\{0,\ \frac{\bar M_1\bar Z-\bar s\,\bar I}
        {\bar s\,\bar M_2-\bar M_1^2}\right\},\\
      \hat b_{0i} &= \frac{\bar Z+\bar M_1\hat b_{1i}}{\bar s}
    \end{aligned}$ \\ \noalign{\medskip}
\begin{tabular}[t]{@{}l@{}}Nonparametric mean\\ $\mu_i\in H(\gamma)$\end{tabular}
  & $\begin{aligned}[t]
      \log f(x_t;\theta_i) &= -\tfrac12\,\lVert x_t-\mu_i\rVert^2\\
      \hat\mu_i &= \frac{\sum_t\iota_t(i)\,x_t}{W_i}
    \end{aligned}$ \\ \noalign{\medskip}
Gaussian on $\mathbb{R}$
  & $\begin{aligned}[t]
      \log f(x_t;\theta_i) &= -\log\sigma_i-\frac{(x_t-\mu_i)^2}{2\sigma_i^2}\\
      \hat\mu_i &= \frac{\sum_t\iota_t(i)\,x_t}{W_i}\\
      \hat\sigma_i^2 &= \frac{\sum_t\iota_t(i)\,(x_t-\hat\mu_i)^2}{W_i}
    \end{aligned}$ \\
\botrule
\end{tabular}
\footnotetext{Curves are observed on a grid $\tau_1<\cdots<\tau_L$, with increments
$\Delta x_{t,k}=x_t(\tau_{k+1})-x_t(\tau_k)$.
\emph{Brownian motion:} $\upsilon=\tfrac12-\nu$,
$\kappa=\Gamma(\upsilon+1)/\Gamma(2\upsilon+1)$, elapsed time
$s_k=\tau_k-\tau_1$,
$S_t=\sum_k \bar g_k\,\Delta x_{t,k}$ with cell averages
$\bar g_k=\kappa\,(s_{k+1}^{\upsilon+1}-s_k^{\upsilon+1})/\{(\upsilon+1)(s_{k+1}-s_k)\}$,
and $Q=\kappa^2 s_L^{2\upsilon+1}/(2\upsilon+1)$; for $\nu=\tfrac12$,
$S_t=x_t(\tau_L)-x_t(\tau_1)$ and $Q=s_L$.
\emph{Ornstein--Uhlenbeck} (left-point sums):
$Z_t=\sum_k\Delta x_{t,k}$,
$I_t=\sum_k x_t(\tau_k)\,\Delta x_{t,k}$,
$M_t^{(1)}=\sum_k x_t(\tau_k)(\tau_{k+1}-\tau_k)$ and
$M_t^{(2)}=\sum_k x_t(\tau_k)^2(\tau_{k+1}-\tau_k)$;
$\bar Z,\bar I,\bar M_1,\bar M_2$ are their $\iota_t(i)$-weighted sums
and $\bar s=W_i s_L$.
\emph{Nonparametric:} the update is exact for the L$^2$, $W_0^{2,1}$ and
$W_0^{2,2}$ norms (unique up to the null space of the latter two
seminorms) and a heuristic for the sup norm.}
\end{table}
\FloatBarrier

\section{Simulation Details}\label{apx:simDetails}
\begingroup
\footnotesize
\setlength{\tabcolsep}{4pt}
\renewcommand{\arraystretch}{0.95}

\begin{longtable}{l l c l r r r r r}
\caption{Geometric truth ($n = 50$ replicates per cell): estimates of matched
states (leave probability prob, negative binomial size and prob, mean dwell,
TV distance between fitted and true dwell pmfs, RMSE of the fitted mean curve),
by separation; $\mathrm{geo}(p) = 1 + \mathrm{NB}(\text{size} = 1, \text{prob} = p)$.
All dwell distributions are geometric. Methods: THMM = THMM (funHMM);
THSMM-G = THSMM, geometric; THSMM-NB = THSMM, negative binomial.}
\label{tab:geomTruth}\\
\toprule
Sep. & Method & State & Parameter & True & Mean & Bias & SD & RMSE \\
\midrule
\endfirsthead

\multicolumn{9}{l}{\tablename~\thetable\ (continued)}\\
\toprule
Sep. & Method & State & Parameter & True & Mean & Bias & SD & RMSE \\
\midrule
\endhead

\midrule
\multicolumn{9}{r}{\emph{continued on next page}}\\
\endfoot

\bottomrule
\endlastfoot

low & THMM & 1 & mean curve RMSE & 0 & 0.148 & 0.148 & 0.0683 & 0.162 \\
low & THMM & 1 & mean dwell & 2.5 & 2.11 & -0.391 & 1.01 & 1.07 \\
low & THMM & 1 & prob & 0.4 & 0.556 & 0.156 & 0.191 & 0.245 \\
low & THMM & 1 & TV dwell & 0 & 0.227 & 0.227 & 0.115 & 0.254 \\
low & THMM & 2 & mean curve RMSE & 0 & 0.0695 & 0.0695 & 0.0283 & 0.0749 \\
low & THMM & 2 & mean dwell & 4 & 5.82 & 1.82 & 2.79 & 3.31 \\
low & THMM & 2 & prob & 0.25 & 0.243 & -0.00728 & 0.167 & 0.165 \\
low & THMM & 2 & TV dwell & 0 & 0.246 & 0.246 & 0.0967 & 0.264 \\
low & THMM & 3 & mean curve RMSE & 0 & 0.0432 & 0.0432 & 0.0215 & 0.0482 \\
low & THMM & 3 & mean dwell & 6.67 & 5.57 & -1.09 & 2.94 & 3.11 \\
low & THMM & 3 & prob & 0.15 & 0.26 & 0.11 & 0.169 & 0.2 \\
low & THMM & 3 & TV dwell & 0 & 0.223 & 0.223 & 0.187 & 0.29 \\
\addlinespace
low & THSMM-G & 1 & mean curve RMSE & 0 & 0.148 & 0.148 & 0.0683 & 0.162 \\
low & THSMM-G & 1 & mean dwell & 2.5 & 2.11 & -0.391 & 1.01 & 1.07 \\
low & THSMM-G & 1 & prob & 0.4 & 0.556 & 0.156 & 0.191 & 0.245 \\
low & THSMM-G & 1 & TV dwell & 0 & 0.227 & 0.227 & 0.115 & 0.254 \\
low & THSMM-G & 2 & mean curve RMSE & 0 & 0.0695 & 0.0695 & 0.0283 & 0.0749 \\
low & THSMM-G & 2 & mean dwell & 4 & 5.82 & 1.82 & 2.79 & 3.31 \\
low & THSMM-G & 2 & prob & 0.25 & 0.243 & -0.00728 & 0.167 & 0.165 \\
low & THSMM-G & 2 & TV dwell & 0 & 0.246 & 0.246 & 0.0967 & 0.264 \\
low & THSMM-G & 3 & mean curve RMSE & 0 & 0.0432 & 0.0432 & 0.0215 & 0.0482 \\
low & THSMM-G & 3 & mean dwell & 6.67 & 5.57 & -1.09 & 2.94 & 3.11 \\
low & THSMM-G & 3 & prob & 0.15 & 0.26 & 0.11 & 0.169 & 0.2 \\
low & THSMM-G & 3 & TV dwell & 0 & 0.223 & 0.223 & 0.187 & 0.29 \\
\addlinespace
low & THSMM-NB & 1 & mean curve RMSE & 0 & 0.148 & 0.148 & 0.0689 & 0.163 \\
low & THSMM-NB & 1 & mean dwell & 2.5 & 2.25 & -0.252 & 1.29 & 1.3 \\
low & THSMM-NB & 1 & prob & 0.4 & 0.574 & 0.174 & 0.196 & 0.261 \\
low & THSMM-NB & 1 & size & 1 & 15.6 & 14.6 & 98.8 & 98.9 \\
low & THSMM-NB & 1 & TV dwell & 0 & 0.268 & 0.268 & 0.122 & 0.294 \\
low & THSMM-NB & 2 & mean curve RMSE & 0 & 0.0677 & 0.0677 & 0.0268 & 0.0727 \\
low & THSMM-NB & 2 & mean dwell & 4 & 6.3 & 2.3 & 3.11 & 3.85 \\
low & THSMM-NB & 2 & prob & 0.25 & 0.379 & 0.129 & 0.21 & 0.244 \\
low & THSMM-NB & 2 & size & 1 & 2.75e+08 & 2.75e+08 & 1.95e+09 & 1.95e+09 \\
low & THSMM-NB & 2 & TV dwell & 0 & 0.378 & 0.378 & 0.131 & 0.4 \\
low & THSMM-NB & 3 & mean curve RMSE & 0 & 0.0447 & 0.0447 & 0.0231 & 0.0502 \\
low & THSMM-NB & 3 & mean dwell & 6.67 & 5.73 & -0.934 & 3.24 & 3.34 \\
low & THSMM-NB & 3 & prob & 0.15 & 0.373 & 0.223 & 0.227 & 0.316 \\
low & THSMM-NB & 3 & size & 1 & 1.35e+11 & 1.35e+11 & 9.54e+11 & 9.54e+11 \\
low & THSMM-NB & 3 & TV dwell & 0 & 0.317 & 0.317 & 0.154 & 0.352 \\
\midrule

medium & THMM & 1 & mean curve RMSE & 0 & 0.0501 & 0.0501 & 0.0864 & 0.0991 \\
medium & THMM & 1 & mean dwell & 2.5 & 2.76 & 0.262 & 0.548 & 0.602 \\
medium & THMM & 1 & prob & 0.4 & 0.383 & -0.0168 & 0.113 & 0.114 \\
medium & THMM & 1 & TV dwell & 0 & 0.0984 & 0.0984 & 0.0774 & 0.125 \\
medium & THMM & 2 & mean curve RMSE & 0 & 0.0241 & 0.0241 & 0.0208 & 0.0317 \\
medium & THMM & 2 & mean dwell & 4 & 5.39 & 1.39 & 0.931 & 1.67 \\
medium & THMM & 2 & prob & 0.25 & 0.19 & -0.0596 & 0.0293 & 0.0663 \\
medium & THMM & 2 & TV dwell & 0 & 0.117 & 0.117 & 0.0639 & 0.133 \\
medium & THMM & 3 & mean curve RMSE & 0 & 0.0176 & 0.0176 & 0.0166 & 0.0241 \\
medium & THMM & 3 & mean dwell & 6.67 & 7.14 & 0.478 & 1.84 & 1.88 \\
medium & THMM & 3 & prob & 0.15 & 0.164 & 0.0142 & 0.102 & 0.102 \\
medium & THMM & 3 & TV dwell & 0 & 0.0968 & 0.0968 & 0.125 & 0.157 \\
\addlinespace
medium & THSMM-G & 1 & mean curve RMSE & 0 & 0.0501 & 0.0501 & 0.0864 & 0.0991 \\
medium & THSMM-G & 1 & mean dwell & 2.5 & 2.76 & 0.262 & 0.548 & 0.602 \\
medium & THSMM-G & 1 & prob & 0.4 & 0.383 & -0.0168 & 0.113 & 0.114 \\
medium & THSMM-G & 1 & TV dwell & 0 & 0.0984 & 0.0984 & 0.0774 & 0.125 \\
medium & THSMM-G & 2 & mean curve RMSE & 0 & 0.0241 & 0.0241 & 0.0208 & 0.0317 \\
medium & THSMM-G & 2 & mean dwell & 4 & 5.39 & 1.39 & 0.931 & 1.67 \\
medium & THSMM-G & 2 & prob & 0.25 & 0.19 & -0.0596 & 0.0293 & 0.0663 \\
medium & THSMM-G & 2 & TV dwell & 0 & 0.117 & 0.117 & 0.0639 & 0.133 \\
medium & THSMM-G & 3 & mean curve RMSE & 0 & 0.0176 & 0.0176 & 0.0166 & 0.0241 \\
medium & THSMM-G & 3 & mean dwell & 6.67 & 7.14 & 0.478 & 1.84 & 1.88 \\
medium & THSMM-G & 3 & prob & 0.15 & 0.164 & 0.0142 & 0.102 & 0.102 \\
medium & THSMM-G & 3 & TV dwell & 0 & 0.0968 & 0.0968 & 0.125 & 0.157 \\
\addlinespace
medium & THSMM-NB & 1 & mean curve RMSE & 0 & 0.05 & 0.05 & 0.0854 & 0.0982 \\
medium & THSMM-NB & 1 & mean dwell & 2.5 & 2.88 & 0.377 & 0.576 & 0.684 \\
medium & THSMM-NB & 1 & prob & 0.4 & 0.492 & 0.0917 & 0.158 & 0.181 \\
medium & THSMM-NB & 1 & size & 1 & 2.18 & 1.18 & 1.72 & 2.07 \\
medium & THSMM-NB & 1 & TV dwell & 0 & 0.162 & 0.162 & 0.0849 & 0.183 \\
medium & THSMM-NB & 2 & mean curve RMSE & 0 & 0.0243 & 0.0243 & 0.0207 & 0.0318 \\
medium & THSMM-NB & 2 & mean dwell & 4 & 5.63 & 1.63 & 0.934 & 1.87 \\
medium & THSMM-NB & 2 & prob & 0.25 & 0.289 & 0.0389 & 0.097 & 0.104 \\
medium & THSMM-NB & 2 & size & 1 & 1.92 & 0.918 & 0.904 & 1.28 \\
medium & THSMM-NB & 2 & TV dwell & 0 & 0.207 & 0.207 & 0.0679 & 0.218 \\
medium & THSMM-NB & 3 & mean curve RMSE & 0 & 0.0179 & 0.0179 & 0.0175 & 0.0249 \\
medium & THSMM-NB & 3 & mean dwell & 6.67 & 7.36 & 0.692 & 1.95 & 2.05 \\
medium & THSMM-NB & 3 & prob & 0.15 & 0.212 & 0.0621 & 0.145 & 0.157 \\
medium & THSMM-NB & 3 & size & 1 & 1.52 & 0.524 & 0.744 & 0.903 \\
medium & THSMM-NB & 3 & TV dwell & 0 & 0.155 & 0.155 & 0.122 & 0.197 \\
\midrule

high & THMM & 1 & mean curve RMSE & 0 & 0.0106 & 0.0106 & 0.00361 & 0.0112 \\
high & THMM & 1 & mean dwell & 2.5 & 2.58 & 0.0802 & 0.246 & 0.257 \\
high & THMM & 1 & prob & 0.4 & 0.391 & -0.00885 & 0.0386 & 0.0393 \\
high & THMM & 1 & TV dwell & 0 & 0.0394 & 0.0394 & 0.0265 & 0.0473 \\
high & THMM & 2 & mean curve RMSE & 0 & 0.00901 & 0.00901 & 0.00351 & 0.00965 \\
high & THMM & 2 & mean dwell & 4 & 4.18 & 0.177 & 0.404 & 0.437 \\
high & THMM & 2 & prob & 0.25 & 0.242 & -0.00831 & 0.0245 & 0.0257 \\
high & THMM & 2 & TV dwell & 0 & 0.0357 & 0.0357 & 0.0267 & 0.0445 \\
high & THMM & 3 & mean curve RMSE & 0 & 0.00688 & 0.00688 & 0.00269 & 0.00737 \\
high & THMM & 3 & mean dwell & 6.67 & 6.7 & 0.029 & 0.786 & 0.779 \\
high & THMM & 3 & prob & 0.15 & 0.151 & 0.0013 & 0.0171 & 0.017 \\
high & THMM & 3 & TV dwell & 0 & 0.0357 & 0.0357 & 0.0281 & 0.0453 \\
\addlinespace
high & THSMM-G & 1 & mean curve RMSE & 0 & 0.0106 & 0.0106 & 0.00361 & 0.0112 \\
high & THSMM-G & 1 & mean dwell & 2.5 & 2.58 & 0.0802 & 0.246 & 0.256 \\
high & THSMM-G & 1 & prob & 0.4 & 0.391 & -0.00885 & 0.0386 & 0.0393 \\
high & THSMM-G & 1 & TV dwell & 0 & 0.0394 & 0.0394 & 0.0265 & 0.0473 \\
high & THSMM-G & 2 & mean curve RMSE & 0 & 0.00901 & 0.00901 & 0.00351 & 0.00965 \\
high & THSMM-G & 2 & mean dwell & 4 & 4.18 & 0.177 & 0.404 & 0.437 \\
high & THSMM-G & 2 & prob & 0.25 & 0.242 & -0.00831 & 0.0245 & 0.0257 \\
high & THSMM-G & 2 & TV dwell & 0 & 0.0357 & 0.0357 & 0.0267 & 0.0445 \\
high & THSMM-G & 3 & mean curve RMSE & 0 & 0.00688 & 0.00688 & 0.00269 & 0.00737 \\
high & THSMM-G & 3 & mean dwell & 6.67 & 6.7 & 0.029 & 0.786 & 0.779 \\
high & THSMM-G & 3 & prob & 0.15 & 0.151 & 0.0013 & 0.0171 & 0.017 \\
high & THSMM-G & 3 & TV dwell & 0 & 0.0357 & 0.0357 & 0.0281 & 0.0453 \\
\addlinespace
high & THSMM-NB & 1 & mean curve RMSE & 0 & 0.0107 & 0.0107 & 0.00364 & 0.0113 \\
high & THSMM-NB & 1 & mean dwell & 2.5 & 2.58 & 0.0848 & 0.249 & 0.261 \\
high & THSMM-NB & 1 & prob & 0.4 & 0.432 & 0.0319 & 0.0791 & 0.0845 \\
high & THSMM-NB & 1 & size & 1 & 1.24 & 0.235 & 0.379 & 0.443 \\
high & THSMM-NB & 1 & TV dwell & 0 & 0.0613 & 0.0613 & 0.033 & 0.0695 \\
high & THSMM-NB & 2 & mean curve RMSE & 0 & 0.00903 & 0.00903 & 0.00352 & 0.00968 \\
high & THSMM-NB & 2 & mean dwell & 4 & 4.19 & 0.186 & 0.407 & 0.444 \\
high & THSMM-NB & 2 & prob & 0.25 & 0.264 & 0.0141 & 0.0539 & 0.0552 \\
high & THSMM-NB & 2 & size & 1 & 1.15 & 0.151 & 0.295 & 0.328 \\
high & THSMM-NB & 2 & TV dwell & 0 & 0.0636 & 0.0636 & 0.0343 & 0.0722 \\
high & THSMM-NB & 3 & mean curve RMSE & 0 & 0.00688 & 0.00688 & 0.00269 & 0.00738 \\
high & THSMM-NB & 3 & mean dwell & 6.67 & 6.7 & 0.0344 & 0.786 & 0.779 \\
high & THSMM-NB & 3 & prob & 0.15 & 0.166 & 0.016 & 0.035 & 0.0382 \\
high & THSMM-NB & 3 & size & 1 & 1.12 & 0.125 & 0.237 & 0.266 \\
high & THSMM-NB & 3 & TV dwell & 0 & 0.0624 & 0.0624 & 0.0263 & 0.0676 \\

\end{longtable}
\endgroup

\begingroup
\footnotesize
\setlength{\tabcolsep}{4pt}
\renewcommand{\arraystretch}{0.95}

\begin{longtable}{l l c l r r r r r}
\caption{Non-geometric truths ($n = 50$ replicates per cell): dwell and
mean-curve estimates of matched states, by separation. Methods: THMM = THMM
(funHMM); THSMM-SP = THSMM, shifted Poisson; THSMM-NB = THSMM, negative
binomial; THSMM-NP = THSMM, nonparametric.}
\label{tab:nongeomTruth}\\
\toprule
Sep. & Method & State & Parameter & True & Mean & Bias & SD & RMSE \\
\midrule
\endfirsthead

\multicolumn{9}{l}{\tablename~\thetable\ (continued)}\\
\toprule
Sep. & Method & State & Parameter & True & Mean & Bias & SD & RMSE \\
\midrule
\endhead

\midrule
\multicolumn{9}{r}{\emph{continued on next page}}\\
\endfoot

\bottomrule
\endlastfoot

\multicolumn{9}{l}{\textbf{True dwell distribution: negative binomial}}\\
\midrule

low & THMM & 1 & mean curve RMSE & 0 & 0.0551 & 0.0551 & 0.0534 & 0.0764 \\
low & THMM & 1 & mean dwell & 4 & 3.69 & -0.314 & 1.11 & 1.15 \\
low & THMM & 1 & TV dwell & 0 & 0.253 & 0.253 & 0.155 & 0.296 \\
low & THMM & 2 & mean curve RMSE & 0 & 0.0358 & 0.0358 & 0.0317 & 0.0476 \\
low & THMM & 2 & mean dwell & 6 & 5.68 & -0.324 & 1.88 & 1.89 \\
low & THMM & 2 & TV dwell & 0 & 0.345 & 0.345 & 0.15 & 0.376 \\
low & THMM & 3 & mean curve RMSE & 0 & 0.0238 & 0.0238 & 0.00783 & 0.0251 \\
low & THMM & 3 & mean dwell & 7.67 & 7.47 & -0.195 & 1.05 & 1.06 \\
low & THMM & 3 & TV dwell & 0 & 0.378 & 0.378 & 0.0431 & 0.381 \\
\addlinespace
low & THSMM-SP & 1 & mean curve RMSE & 0 & 0.0321 & 0.0321 & 0.0182 & 0.0368 \\
low & THSMM-SP & 1 & mean dwell & 4 & 4.67 & 0.668 & 0.726 & 0.981 \\
low & THSMM-SP & 1 & TV dwell & 0 & 0.264 & 0.264 & 0.075 & 0.274 \\
low & THSMM-SP & 2 & mean curve RMSE & 0 & 0.019 & 0.019 & 0.00783 & 0.0205 \\
low & THSMM-SP & 2 & mean dwell & 6 & 6.93 & 0.926 & 0.832 & 1.24 \\
low & THSMM-SP & 2 & TV dwell & 0 & 0.262 & 0.262 & 0.0746 & 0.272 \\
low & THSMM-SP & 3 & mean curve RMSE & 0 & 0.0182 & 0.0182 & 0.0045 & 0.0188 \\
low & THSMM-SP & 3 & mean dwell & 7.67 & 8.41 & 0.743 & 0.845 & 1.12 \\
low & THSMM-SP & 3 & TV dwell & 0 & 0.194 & 0.194 & 0.066 & 0.205 \\
\addlinespace
low & THSMM-NB & 1 & mean curve RMSE & 0 & 0.0499 & 0.0499 & 0.0527 & 0.0722 \\
low & THSMM-NB & 1 & mean dwell & 4 & 4.12 & 0.117 & 1.26 & 1.26 \\
low & THSMM-NB & 1 & prob & 0.5 & 0.646 & 0.146 & 0.174 & 0.226 \\
low & THSMM-NB & 1 & size & 3 & 9.55 & 6.55 & 11.5 & 13.1 \\
low & THSMM-NB & 1 & TV dwell & 0 & 0.228 & 0.228 & 0.167 & 0.282 \\
low & THSMM-NB & 2 & mean curve RMSE & 0 & 0.0315 & 0.0315 & 0.031 & 0.044 \\
low & THSMM-NB & 2 & mean dwell & 6 & 6.3 & 0.299 & 2.03 & 2.03 \\
low & THSMM-NB & 2 & prob & 0.5 & 0.634 & 0.134 & 0.193 & 0.233 \\
low & THSMM-NB & 2 & size & 5 & 1.19e+11 & 1.19e+11 & 5.18e+11 & 5.27e+11 \\
low & THSMM-NB & 2 & TV dwell & 0 & 0.247 & 0.247 & 0.199 & 0.316 \\
low & THSMM-NB & 3 & mean curve RMSE & 0 & 0.0195 & 0.0195 & 0.00866 & 0.0213 \\
low & THSMM-NB & 3 & mean dwell & 7.67 & 8.27 & 0.6 & 1.11 & 1.25 \\
low & THSMM-NB & 3 & prob & 0.6 & 0.677 & 0.0769 & 0.227 & 0.237 \\
low & THSMM-NB & 3 & size & 10 & 9.12e+10 & 9.12e+10 & 4.09e+11 & 4.15e+11 \\
low & THSMM-NB & 3 & TV dwell & 0 & 0.133 & 0.133 & 0.103 & 0.168 \\
\addlinespace
low & THSMM-NP & 1 & mean curve RMSE & 0 & 0.0402 & 0.0402 & 0.0517 & 0.0651 \\
low & THSMM-NP & 1 & mean dwell & 4 & 4.7 & 0.704 & 1.03 & 1.24 \\
low & THSMM-NP & 1 & TV dwell & 0 & 0.525 & 0.525 & 0.111 & 0.536 \\
low & THSMM-NP & 2 & mean curve RMSE & 0 & 0.0229 & 0.0229 & 0.0214 & 0.0312 \\
low & THSMM-NP & 2 & mean dwell & 6 & 7.09 & 1.09 & 1.34 & 1.71 \\
low & THSMM-NP & 2 & TV dwell & 0 & 0.555 & 0.555 & 0.0889 & 0.562 \\
low & THSMM-NP & 3 & mean curve RMSE & 0 & 0.018 & 0.018 & 0.00469 & 0.0185 \\
low & THSMM-NP & 3 & mean dwell & 7.67 & 8.66 & 0.989 & 0.673 & 1.19 \\
low & THSMM-NP & 3 & TV dwell & 0 & 0.485 & 0.485 & 0.0751 & 0.49 \\
\midrule

medium & THMM & 1 & mean curve RMSE & 0 & 0.0131 & 0.0131 & 0.00454 & 0.0138 \\
medium & THMM & 1 & mean dwell & 4 & 4.14 & 0.145 & 0.335 & 0.362 \\
medium & THMM & 1 & TV dwell & 0 & 0.177 & 0.177 & 0.006 & 0.177 \\
medium & THMM & 2 & mean curve RMSE & 0 & 0.011 & 0.011 & 0.00343 & 0.0115 \\
medium & THMM & 2 & mean dwell & 6 & 6.04 & 0.0357 & 0.449 & 0.446 \\
medium & THMM & 2 & TV dwell & 0 & 0.274 & 0.274 & 0.00312 & 0.274 \\
medium & THMM & 3 & mean curve RMSE & 0 & 0.0109 & 0.0109 & 0.00377 & 0.0115 \\
medium & THMM & 3 & mean dwell & 7.67 & 7.77 & 0.108 & 0.457 & 0.465 \\
medium & THMM & 3 & TV dwell & 0 & 0.369 & 0.369 & 0.0022 & 0.369 \\
\addlinespace
medium & THSMM-SP & 1 & mean curve RMSE & 0 & 0.0138 & 0.0138 & 0.00485 & 0.0146 \\
medium & THSMM-SP & 1 & mean dwell & 4 & 4.38 & 0.378 & 0.389 & 0.539 \\
medium & THSMM-SP & 1 & TV dwell & 0 & 0.205 & 0.205 & 0.0358 & 0.208 \\
medium & THSMM-SP & 2 & mean curve RMSE & 0 & 0.0109 & 0.0109 & 0.00313 & 0.0113 \\
medium & THSMM-SP & 2 & mean dwell & 6 & 6.32 & 0.324 & 0.483 & 0.577 \\
medium & THSMM-SP & 2 & TV dwell & 0 & 0.191 & 0.191 & 0.0301 & 0.193 \\
medium & THSMM-SP & 3 & mean curve RMSE & 0 & 0.0101 & 0.0101 & 0.00363 & 0.0107 \\
medium & THSMM-SP & 3 & mean dwell & 7.67 & 7.97 & 0.308 & 0.438 & 0.532 \\
medium & THSMM-SP & 3 & TV dwell & 0 & 0.144 & 0.144 & 0.0245 & 0.146 \\
\addlinespace
medium & THSMM-NB & 1 & mean curve RMSE & 0 & 0.0128 & 0.0128 & 0.00438 & 0.0135 \\
medium & THSMM-NB & 1 & mean dwell & 4 & 4.36 & 0.36 & 0.39 & 0.528 \\
medium & THSMM-NB & 1 & prob & 0.5 & 0.622 & 0.122 & 0.117 & 0.169 \\
medium & THSMM-NB & 1 & size & 3 & 6.52 & 3.52 & 3.88 & 5.21 \\
medium & THSMM-NB & 1 & TV dwell & 0 & 0.106 & 0.106 & 0.0494 & 0.116 \\
medium & THSMM-NB & 2 & mean curve RMSE & 0 & 0.0104 & 0.0104 & 0.0031 & 0.0109 \\
medium & THSMM-NB & 2 & mean dwell & 6 & 6.33 & 0.335 & 0.463 & 0.568 \\
medium & THSMM-NB & 2 & prob & 0.5 & 0.567 & 0.0673 & 0.136 & 0.15 \\
medium & THSMM-NB & 2 & size & 5 & 9.29 & 4.29 & 8.98 & 9.88 \\
medium & THSMM-NB & 2 & TV dwell & 0 & 0.0814 & 0.0814 & 0.0478 & 0.0942 \\
medium & THSMM-NB & 3 & mean curve RMSE & 0 & 0.00996 & 0.00996 & 0.00361 & 0.0106 \\
medium & THSMM-NB & 3 & mean dwell & 7.67 & 8 & 0.338 & 0.456 & 0.564 \\
medium & THSMM-NB & 3 & prob & 0.6 & 0.662 & 0.0622 & 0.132 & 0.145 \\
medium & THSMM-NB & 3 & size & 10 & 2.99e+07 & 2.99e+07 & 2.11e+08 & 2.11e+08 \\
medium & THSMM-NB & 3 & TV dwell & 0 & 0.0742 & 0.0742 & 0.035 & 0.0819 \\
\addlinespace
medium & THSMM-NP & 1 & mean curve RMSE & 0 & 0.013 & 0.013 & 0.00414 & 0.0137 \\
medium & THSMM-NP & 1 & mean dwell & 4 & 4.5 & 0.498 & 0.374 & 0.621 \\
medium & THSMM-NP & 1 & TV dwell & 0 & 0.277 & 0.277 & 0.0647 & 0.285 \\
medium & THSMM-NP & 2 & mean curve RMSE & 0 & 0.0106 & 0.0106 & 0.00307 & 0.011 \\
medium & THSMM-NP & 2 & mean dwell & 6 & 6.44 & 0.443 & 0.486 & 0.653 \\
medium & THSMM-NP & 2 & TV dwell & 0 & 0.296 & 0.296 & 0.0634 & 0.302 \\
medium & THSMM-NP & 3 & mean curve RMSE & 0 & 0.00986 & 0.00986 & 0.00362 & 0.0105 \\
medium & THSMM-NP & 3 & mean dwell & 7.67 & 8.08 & 0.409 & 0.48 & 0.627 \\
medium & THSMM-NP & 3 & TV dwell & 0 & 0.268 & 0.268 & 0.0602 & 0.274 \\
\midrule

high & THMM & 1 & mean curve RMSE & 0 & 0.00932 & 0.00932 & 0.00431 & 0.0103 \\
high & THMM & 1 & mean dwell & 4 & 4.04 & 0.0386 & 0.338 & 0.337 \\
high & THMM & 1 & TV dwell & 0 & 0.177 & 0.177 & 0.00662 & 0.177 \\
high & THMM & 2 & mean curve RMSE & 0 & 0.00745 & 0.00745 & 0.00257 & 0.00788 \\
high & THMM & 2 & mean dwell & 6 & 6.14 & 0.135 & 0.505 & 0.518 \\
high & THMM & 2 & TV dwell & 0 & 0.275 & 0.275 & 0.00549 & 0.275 \\
high & THMM & 3 & mean curve RMSE & 0 & 0.00647 & 0.00647 & 0.00348 & 0.00733 \\
high & THMM & 3 & mean dwell & 7.67 & 7.72 & 0.0549 & 0.39 & 0.39 \\
high & THMM & 3 & TV dwell & 0 & 0.368 & 0.368 & 0.00179 & 0.368 \\
\addlinespace
high & THSMM-SP & 1 & mean curve RMSE & 0 & 0.00952 & 0.00952 & 0.00427 & 0.0104 \\
high & THSMM-SP & 1 & mean dwell & 4 & 4.09 & 0.0915 & 0.36 & 0.368 \\
high & THSMM-SP & 1 & TV dwell & 0 & 0.183 & 0.183 & 0.0279 & 0.185 \\
high & THSMM-SP & 2 & mean curve RMSE & 0 & 0.00741 & 0.00741 & 0.00261 & 0.00785 \\
high & THSMM-SP & 2 & mean dwell & 6 & 6.19 & 0.188 & 0.522 & 0.55 \\
high & THSMM-SP & 2 & TV dwell & 0 & 0.186 & 0.186 & 0.0299 & 0.188 \\
high & THSMM-SP & 3 & mean curve RMSE & 0 & 0.00653 & 0.00653 & 0.00347 & 0.00737 \\
high & THSMM-SP & 3 & mean dwell n& 7.67 & 7.73 & 0.0646 & 0.393 & 0.395 \\
high & THSMM-SP & 3 & TV dwell & 0 & 0.133 & 0.133 & 0.0122 & 0.134 \\
\addlinespace
high & THSMM-NB & 1 & mean curve RMSE & 0 & 0.00938 & 0.00938 & 0.00429 & 0.0103 \\
high & THSMM-NB & 1 & mean dwell & 4 & 4.07 & 0.0709 & 0.346 & 0.349 \\
high & THSMM-NB & 1 & prob & 0.5 & 0.542 & 0.0418 & 0.107 & 0.113 \\
high & THSMM-NB & 1 & size & 3 & 4.11 & 1.11 & 2.4 & 2.62 \\
high & THSMM-NB & 1 & TV dwell & 0 & 0.0649 & 0.0649 & 0.0358 & 0.0739 \\
high & THSMM-NB & 2 & mean curve RMSE & 0 & 0.00742 & 0.00742 & 0.00258 & 0.00785 \\
high & THSMM-NB & 2 & mean dwell & 6 & 6.17 & 0.172 & 0.511 & 0.535 \\
high & THSMM-NB & 2 & prob & 0.5 & 0.539 & 0.0386 & 0.109 & 0.115 \\
high & THSMM-NB & 2 & size & 5 & 6.75 & 1.75 & 3.48 & 3.87 \\
high & THSMM-NB & 2 & TV dwell & 0 & 0.0677 & 0.0677 & 0.0425 & 0.0797 \\
high & THSMM-NB & 3 & mean curve RMSE & 0 & 0.0065 & 0.0065 & 0.00347 & 0.00735 \\
high & THSMM-NB & 3 & mean dwell & 7.67 & 7.73 & 0.0606 & 0.387 & 0.388 \\
high & THSMM-NB & 3 & prob & 0.6 & 0.621 & 0.0208 & 0.133 & 0.133 \\
high & THSMM-NB & 3 & size & 10 & 20.5 & 10.5 & 40.1 & 41.1 \\
high & THSMM-NB & 3 & TV dwell & 0 & 0.0585 & 0.0585 & 0.0318 & 0.0664 \\
\addlinespace
high & THSMM-NP & 1 & mean curve RMSE & 0 & 0.0094 & 0.0094 & 0.00438 & 0.0104 \\
high & THSMM-NP & 1 & mean dwell & 4 & 4.07 & 0.0723 & 0.36 & 0.364 \\
high & THSMM-NP & 1 & TV dwell & 0 & 0.17 & 0.17 & 0.0468 & 0.176 \\
high & THSMM-NP & 2 & mean curve RMSE & 0 & 0.0074 & 0.0074 & 0.00259 & 0.00783 \\
high & THSMM-NP & 2 & mean dwell & 6 & 6.18 & 0.18 & 0.521 & 0.547 \\
high & THSMM-NP & 2 & TV dwell & 0 & 0.203 & 0.203 & 0.0462 & 0.208 \\
high & THSMM-NP & 3 & mean curve RMSE & 0 & 0.00649 & 0.00649 & 0.00347 & 0.00734 \\
high & THSMM-NP & 3 & mean dwell & 7.67 & 7.72 & 0.0563 & 0.383 & 0.383 \\
high & THSMM-NP & 3 & TV dwell & 0 & 0.193 & 0.193 & 0.0402 & 0.197 \\
\midrule

\multicolumn{9}{l}{\textbf{True dwell distribution: shifted Poisson}}\\
\midrule

low & THMM & 1 & mean curve RMSE & 0 & 0.0398 & 0.0398 & 0.0214 & 0.0451 \\
low & THMM & 1 & mean dwell & 4 & 3.25 & -0.748 & 0.463 & 0.877 \\
low & THMM & 1 & TV dwell & 0 & 0.622 & 0.622 & 0.0451 & 0.624 \\
low & THMM & 2 & mean curve RMSE & 0 & 0.0565 & 0.0565 & 0.024 & 0.0613 \\
low & THMM & 2 & mean dwell & 4 & 2.96 & -1.04 & 0.542 & 1.17 \\
low & THMM & 2 & TV dwell & 0 & 0.484 & 0.484 & 0.0832 & 0.491 \\
low & THMM & 3 & mean curve RMSE & 0 & 0.0507 & 0.0507 & 0.00904 & 0.0515 \\
low & THMM & 3 & mean dwell & 5 & 3.68 & -1.32 & 0.468 & 1.4 \\
low & THMM & 3 & TV dwell & 0 & 0.709 & 0.709 & 0.0251 & 0.709 \\
\addlinespace
low & THSMM-SP & 1 & $\lambda$ & 1 & 1.07 & 0.0652 & 0.133 & 0.147 \\
low & THSMM-SP & 1 & mean curve RMSE & 0 & 0.015 & 0.015 & 0.00485 & 0.0158 \\
low & THSMM-SP & 1 & mean dwell & 4 & 4.09 & 0.0852 & 0.152 & 0.173 \\
low & THSMM-SP & 1 & shift & 3 & 3.02 & 0.02 & 0.141 & 0.141 \\
low & THSMM-SP & 1 & TV dwell & 0 & 0.0499 & 0.0499 & 0.0547 & 0.0737 \\
low & THSMM-SP & 2 & $\lambda$ & 2 & 1.45 & -0.55 & 0.462 & 0.715 \\
low & THSMM-SP & 2 & mean curve RMSE & 0 & 0.0188 & 0.0188 & 0.00581 & 0.0197 \\
low & THSMM-SP & 2 & mean dwell & 4 & 4.07 & 0.0701 & 0.21 & 0.22 \\
low & THSMM-SP & 2 & shift & 2 & 2.62 & 0.62 & 0.49 & 0.787 \\
low & THSMM-SP & 2 & TV dwell & 0 & 0.134 & 0.134 & 0.0766 & 0.154 \\
low & THSMM-SP & 3 & $\lambda$ & 1 & 1.06 & 0.0648 & 0.183 & 0.193 \\
low & THSMM-SP & 3 & mean curve RMSE & 0 & 0.0165 & 0.0165 & 0.00455 & 0.0171 \\
low & THSMM-SP & 3 & mean dwell & 5 & 5.04 & 0.0448 & 0.139 & 0.145 \\
low & THSMM-SP & 3 & shift & 4 & 3.98 & -0.02 & 0.141 & 0.141 \\
low & THSMM-SP & 3 & TV dwell & 0 & 0.0441 & 0.0441 & 0.0404 & 0.0595 \\
\addlinespace
low & THSMM-NB & 1 & mean curve RMSE & 0 & 0.0264 & 0.0264 & 0.0174 & 0.0315 \\
low & THSMM-NB & 1 & mean dwell & 4 & 2.76 & -1.24 & 0.376 & 1.29 \\
low & THSMM-NB & 1 & TV dwell & 0 & 0.517 & 0.517 & 0.085 & 0.523 \\
low & THSMM-NB & 2 & mean curve RMSE & 0 & 0.0384 & 0.0384 & 0.0194 & 0.0429 \\
low & THSMM-NB & 2 & mean dwell & 4 & 3.14 & -0.865 & 0.495 & 0.994 \\
low & THSMM-NB & 2 & TV dwell & 0 & 0.299 & 0.299 & 0.125 & 0.324 \\
low & THSMM-NB & 3 & mean curve RMSE & 0 & 0.0363 & 0.0363 & 0.011 & 0.0379 \\
low & THSMM-NB & 3 & mean dwell & 5 & 4.05 & -0.95 & 0.543 & 1.09 \\
low & THSMM-NB & 3 & TV dwell & 0 & 0.489 & 0.489 & 0.0883 & 0.497 \\
\addlinespace
low & THSMM-NP & 1 & mean curve RMSE & 0 & 0.0147 & 0.0147 & 0.00483 & 0.0155 \\
low & THSMM-NP & 1 & mean dwell & 4 & 4.08 & 0.0787 & 0.148 & 0.166 \\
low & THSMM-NP & 1 & TV dwell & 0 & 0.224 & 0.224 & 0.107 & 0.248 \\
low & THSMM-NP & 2 & mean curve RMSE & 0 & 0.019 & 0.019 & 0.00599 & 0.0199 \\
low & THSMM-NP & 2 & mean dwell & 4 & 4.1 & 0.098 & 0.202 & 0.223 \\
low & THSMM-NP & 2 & TV dwell & 0 & 0.277 & 0.277 & 0.0954 & 0.292 \\
low & THSMM-NP & 3 & mean curve RMSE & 0 & 0.0164 & 0.0164 & 0.00475 & 0.0171 \\
low & THSMM-NP & 3 & mean dwell & 5 & 5.02 & 0.0221 & 0.19 & 0.19 \\
low & THSMM-NP & 3 & TV dwell & 0 & 0.169 & 0.169 & 0.0781 & 0.186 \\
\midrule

medium & THMM & 1 & mean curve RMSE & 0 & 0.0133 & 0.0133 & 0.00456 & 0.014 \\
medium & THMM & 1 & mean dwell & 4 & 3.75 & -0.254 & 0.133 & 0.286 \\
medium & THMM & 1 & TV dwell & 0 & 0.599 & 0.599 & 0.00187 & 0.599 \\
medium & THMM & 2 & mean curve RMSE & 0 & 0.0157 & 0.0157 & 0.00532 & 0.0166 \\
medium & THMM & 2 & mean dwell & 4 & 3.75 & -0.248 & 0.186 & 0.309 \\
medium & THMM & 2 & TV dwell & 0 & 0.43 & 0.43 & 0.00284 & 0.43 \\
medium & THMM & 3 & mean curve RMSE & 0 & 0.0153 & 0.0153 & 0.00434 & 0.0159 \\
medium & THMM & 3 & mean dwell & 5 & 4.73 & -0.274 & 0.147 & 0.31 \\
medium & THMM & 3 & TV dwell & 0 & 0.681 & 0.681 & 0.00127 & 0.681 \\
\addlinespace
medium & THSMM-SP & 1 & $\lambda$ & 1 & 2.05 & 1.05 & 0.192 & 1.06 \\
medium & THSMM-SP & 1 & mean curve RMSE & 0 & 0.0101 & 0.0101 & 0.00381 & 0.0107 \\
medium & THSMM-SP & 1 & mean dwell & 4 & 4.15 & 0.147 & 0.235 & 0.275 \\
medium & THSMM-SP & 1 & shift & 3 & 2.1 & -0.9 & 0.303 & 0.949 \\
medium & THSMM-SP & 1 & TV dwell & 0 & 0.213 & 0.213 & 0.0231 & 0.215 \\
medium & THSMM-SP & 2 & $\lambda$ & 2 & 2.07 & 0.0665 & 0.254 & 0.26 \\
medium & THSMM-SP & 2 & mean curve RMSE & 0 & 0.0113 & 0.0113 & 0.00477 & 0.0123 \\
medium & THSMM-SP & 2 & mean dwell & 4 & 4.05 & 0.0465 & 0.173 & 0.178 \\
medium & THSMM-SP & 2 & shift & 2 & 1.98 & -0.02 & 0.141 & 0.141 \\
medium & THSMM-SP & 2 & TV dwell & 0 & 0.0409 & 0.0409 & 0.031 & 0.0511 \\
medium & THSMM-SP & 3 & $\lambda$ & 1 & 2.45 & 1.45 & 0.337 & 1.48 \\
medium & THSMM-SP & 3 & mean curve RMSE & 0 & 0.0102 & 0.0102 & 0.00346 & 0.0108 \\
medium & THSMM-SP & 3 & mean dwell & 5 & 5.33 & 0.327 & 0.266 & 0.42 \\
medium & THSMM-SP & 3 & shift & 4 & 2.88 & -1.12 & 0.328 & 1.17 \\
medium & THSMM-SP & 3 & TV dwell & 0 & 0.26 & 0.26 & 0.0407 & 0.263 \\
\addlinespace
medium & THSMM-NB & 1 & mean curve RMSE & 0 & 0.0113 & 0.0113 & 0.00416 & 0.0121 \\
medium & THSMM-NB & 1 & mean dwell & 4 & 3.2 & -0.801 & 0.211 & 0.828 \\
medium & THSMM-NB & 1 & TV dwell & 0 & 0.43 & 0.43 & 0.0365 & 0.431 \\
medium & THSMM-NB & 2 & mean curve RMSE & 0 & 0.0134 & 0.0134 & 0.00507 & 0.0143 \\
medium & THSMM-NB & 2 & mean dwell & 4 & 3.2 & -0.801 & 0.327 & 0.864 \\
medium & THSMM-NB & 2 & TV dwell & 0 & 0.274 & 0.274 & 0.0619 & 0.281 \\
medium & THSMM-NB & 3 & mean curve RMSE & 0 & 0.0122 & 0.0122 & 0.00402 & 0.0128 \\
medium & THSMM-NB & 3 & mean dwell & 5 & 4.1 & -0.9 & 0.359 & 0.967 \\
medium & THSMM-NB & 3 & TV dwell & 0 & 0.474 & 0.474 & 0.0545 & 0.477 \\
\addlinespace
medium & THSMM-NP & 1 & mean curve RMSE & 0 & 0.00946 & 0.00946 & 0.00361 & 0.0101 \\
medium & THSMM-NP & 1 & mean dwell & 4 & 3.99 & -0.0102 & 0.121 & 0.12 \\
medium & THSMM-NP & 1 & TV dwell & 0 & 0.12 & 0.12 & 0.0519 & 0.131 \\
medium & THSMM-NP & 2 & mean curve RMSE & 0 & 0.0115 & 0.0115 & 0.00442 & 0.0123 \\
medium & THSMM-NP & 2 & mean dwell & 4 & 4.05 & 0.0508 & 0.183 & 0.188 \\
medium & THSMM-NP & 2 & TV dwell & 0 & 0.137 & 0.137 & 0.0497 & 0.145 \\
medium & THSMM-NP & 3 & mean curve RMSE & 0 & 0.00912 & 0.00912 & 0.00351 & 0.00975 \\
medium & THSMM-NP & 3 & mean dwell & 5 & 4.97 & -0.0316 & 0.12 & 0.123 \\
medium & THSMM-NP & 3 & TV dwell & 0 & 0.102 & 0.102 & 0.0409 & 0.11 \\
\midrule

high & THMM & 1 & mean curve RMSE & 0 & 0.00897 & 0.00897 & 0.0036 & 0.00965 \\
high & THMM & 1 & mean dwell & 4 & 4 & 0.001 & 0.101 & 0.1 \\
high & THMM & 1 & TV dwell & 0 & 0.597 & 0.597 & 0.000488 & 0.597 \\
high & THMM & 2 & mean curve RMSE & 0 & 0.00879 & 0.00879 & 0.00361 & 0.00949 \\
high & THMM & 2 & mean dwell & 4 & 3.95 & -0.0543 & 0.152 & 0.16 \\
high & THMM & 2 & TV dwell & 0 & 0.428 & 0.428 & 0.00121 & 0.428 \\
high & THMM & 3 & mean curve RMSE & 0 & 0.00758 & 0.00758 & 0.00326 & 0.00824 \\
high & THMM & 3 & mean dwell & 5 & 4.98 & -0.0168 & 0.106 & 0.107 \\
high & THMM & 3 & TV dwell & 0 & 0.679 & 0.679 & 0.000393 & 0.679 \\
\addlinespace
high & THSMM-SP & 1 & $\lambda$ & 1 & 2.84 & 1.84 & 0.186 & 1.85 \\
high & THSMM-SP & 1 & mean curve RMSE & 0 & 0.00899 & 0.00899 & 0.00359 & 0.00967 \\
high & THSMM-SP & 1 & mean dwell & 4 & 4.44 & 0.442 & 0.367 & 0.572 \\
high & THSMM-SP & 1 & shift & 3 & 1.6 & -1.4 & 0.495 & 1.48 \\
high & THSMM-SP & 1 & TV dwell & 0 & 0.31 & 0.31 & 0.0205 & 0.31 \\
high & THSMM-SP & 2 & $\lambda$ & 2 & 2.96 & 0.96 & 0.153 & 0.972 \\
high & THSMM-SP & 2 & mean curve RMSE & 0 & 0.00872 & 0.00872 & 0.00357 & 0.00941 \\
high & THSMM-SP & 2 & mean dwell & 4 & 3.96 & -0.0396 & 0.153 & 0.157 \\
high & THSMM-SP & 2 & shift & 2 & 1 & -1 & 0 & 1 \\
high & THSMM-SP & 2 & TV dwell & 0 & 0.109 & 0.109 & 0.00695 & 0.109 \\
high & THSMM-SP & 3 & $\lambda$ & 1 & 3.76 & 2.76 & 0.193 & 2.76 \\
high & THSMM-SP & 3 & mean curve RMSE & 0 & 0.00754 & 0.00754 & 0.00325 & 0.0082 \\
high & THSMM-SP & 3 & mean dwell & 5 & 5.58 & 0.576 & 0.307 & 0.651 \\
high & THSMM-SP & 3 & shift & 4 & 1.82 & -2.18 & 0.388 & 2.21 \\
high & THSMM-SP & 3 & TV dwell & 0 & 0.361 & 0.361 & 0.0186 & 0.361 \\
\addlinespace
high & THSMM-NB & 1 & mean curve RMSE & 0 & 0.00899 & 0.00899 & 0.00359 & 0.00967 \\
high & THSMM-NB & 1 & mean dwell & 4 & 3.91 & -0.0899 & 0.0995 & 0.133 \\
high & THSMM-NB & 1 & TV dwell & 0 & 0.357 & 0.357 & 0.00323 & 0.357 \\
high & THSMM-NB & 2 & mean curve RMSE & 0 & 0.00874 & 0.00874 & 0.00357 & 0.00943 \\
high & THSMM-NB & 2 & mean dwell & 4 & 3.89 & -0.11 & 0.143 & 0.179 \\
high & THSMM-NB & 2 & TV dwell & 0 & 0.164 & 0.164 & 0.0101 & 0.164 \\
high & THSMM-NB & 3 & mean curve RMSE & 0 & 0.00755 & 0.00755 & 0.00325 & 0.0082 \\
high & THSMM-NB & 3 & mean dwell & 5 & 4.76 & -0.241 & 0.0938 & 0.258 \\
high & THSMM-NB & 3 & TV dwell & 0 & 0.408 & 0.408 & 0.00389 & 0.408 \\
\addlinespace
high & THSMM-NP & 1 & mean curve RMSE & 0 & 0.00901 & 0.00901 & 0.00356 & 0.00967 \\
high & THSMM-NP & 1 & mean dwell & 4 & 4.01 & 0.0144 & 0.101 & 0.101 \\
high & THSMM-NP & 1 & TV dwell & 0 & 0.0835 & 0.0835 & 0.0318 & 0.0892 \\
high & THSMM-NP & 2 & mean curve RMSE & 0 & 0.00876 & 0.00876 & 0.00357 & 0.00944 \\
high & THSMM-NP & 2 & mean dwell & 4 & 3.97 & -0.0332 & 0.154 & 0.156 \\
high & THSMM-NP & 2 & TV dwell & 0 & 0.097 & 0.097 & 0.0317 & 0.102 \\
high & THSMM-NP & 3 & mean curve RMSE & 0 & 0.00752 & 0.00752 & 0.00326 & 0.00819 \\
high & THSMM-NP & 3 & mean dwell & 5 & 4.99 & -0.0123 & 0.105 & 0.105 \\
high & THSMM-NP & 3 & TV dwell & 0 & 0.0805 & 0.0805 & 0.0299 & 0.0858 \\

\end{longtable}
\endgroup

\begin{figure}[H]
    \centering
    \includegraphics[width=0.8\linewidth]{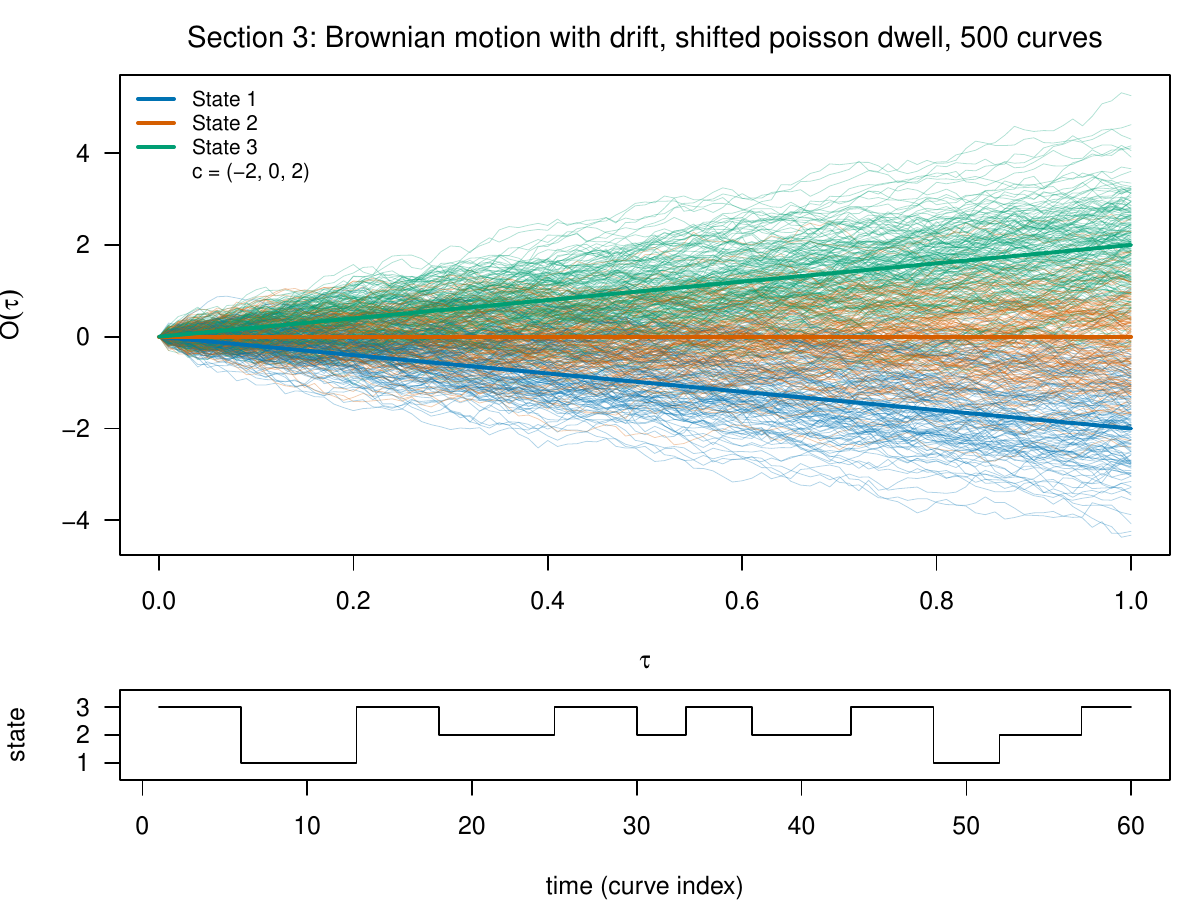}
    \caption{Simulated Brownian motion with drift form a three state semi-Markov process where sojourn distributions are shifted Poisson.}
    \label{fig:bmPois}
\end{figure}

\begin{figure}[H]
    \centering
    \includegraphics[width=0.8\linewidth]{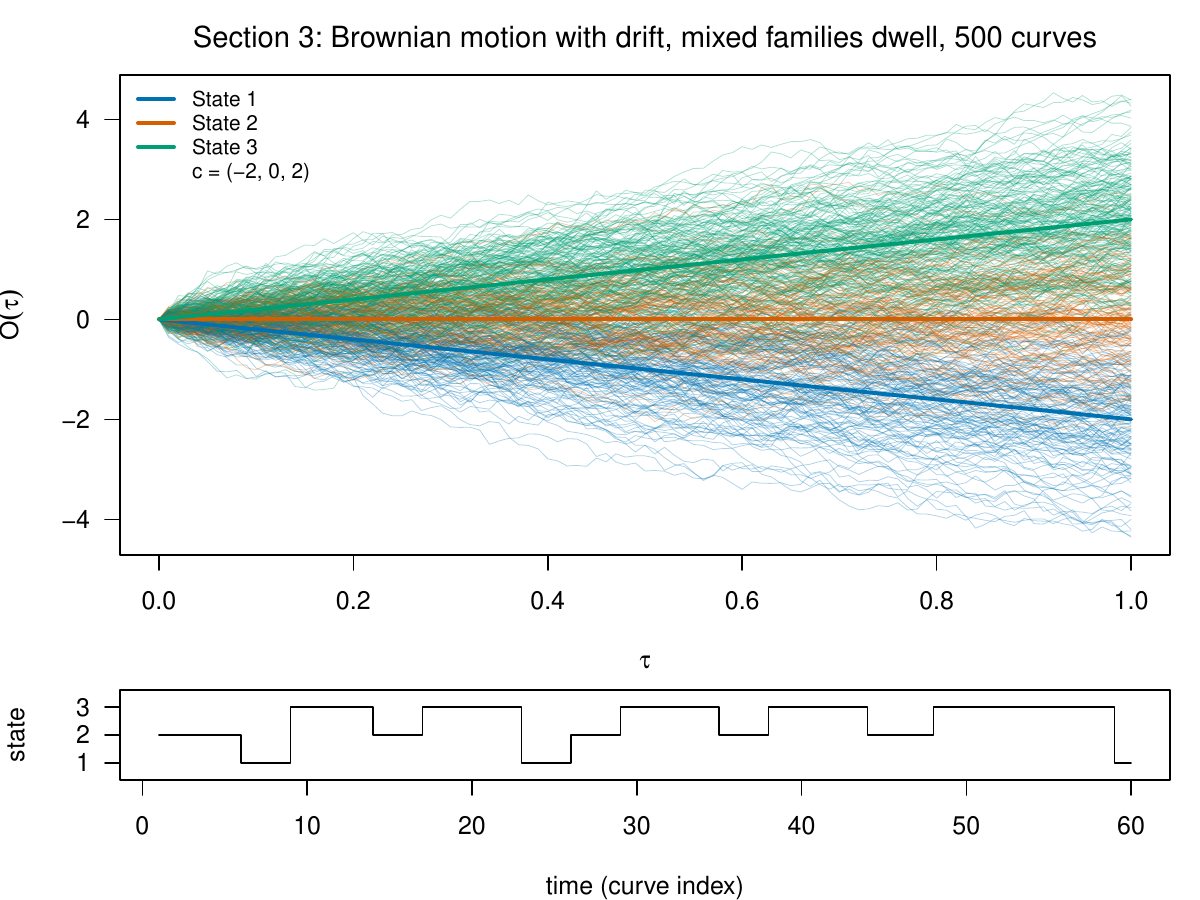}
    \caption{Simulated Brownian motion with drift form a three state semi-Markov process where sojourn distributions are a mixture of families.}
    \label{fig:bmMix}
\end{figure}

\begin{figure}[H]
    \centering
    \includegraphics[width=0.8\linewidth]{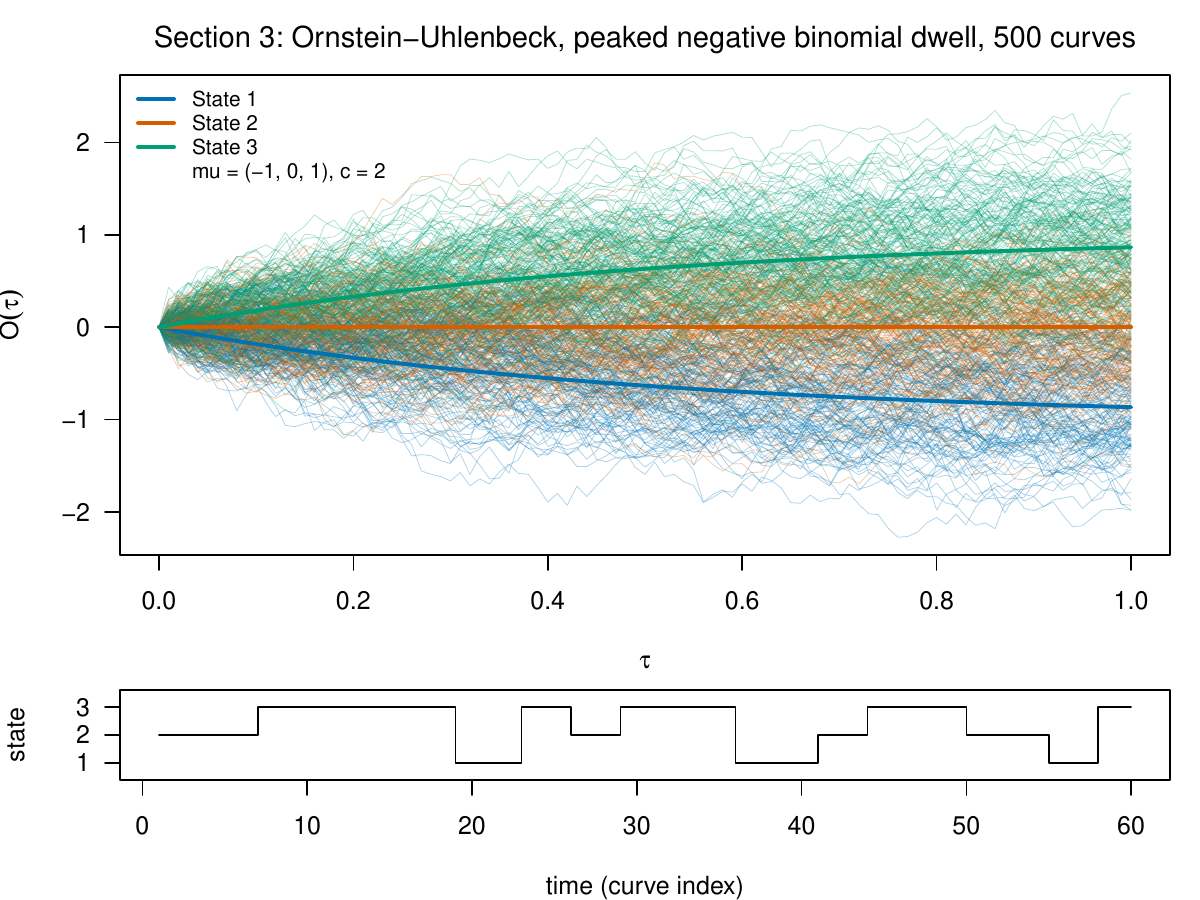}
    \caption{Simulated OU from a three state semi-Markov process where sojourn distributions are negative binomial.}
    \label{fig:ouNb}
\end{figure}

\begin{figure}[H]
    \centering
    \includegraphics[width=0.8\linewidth]{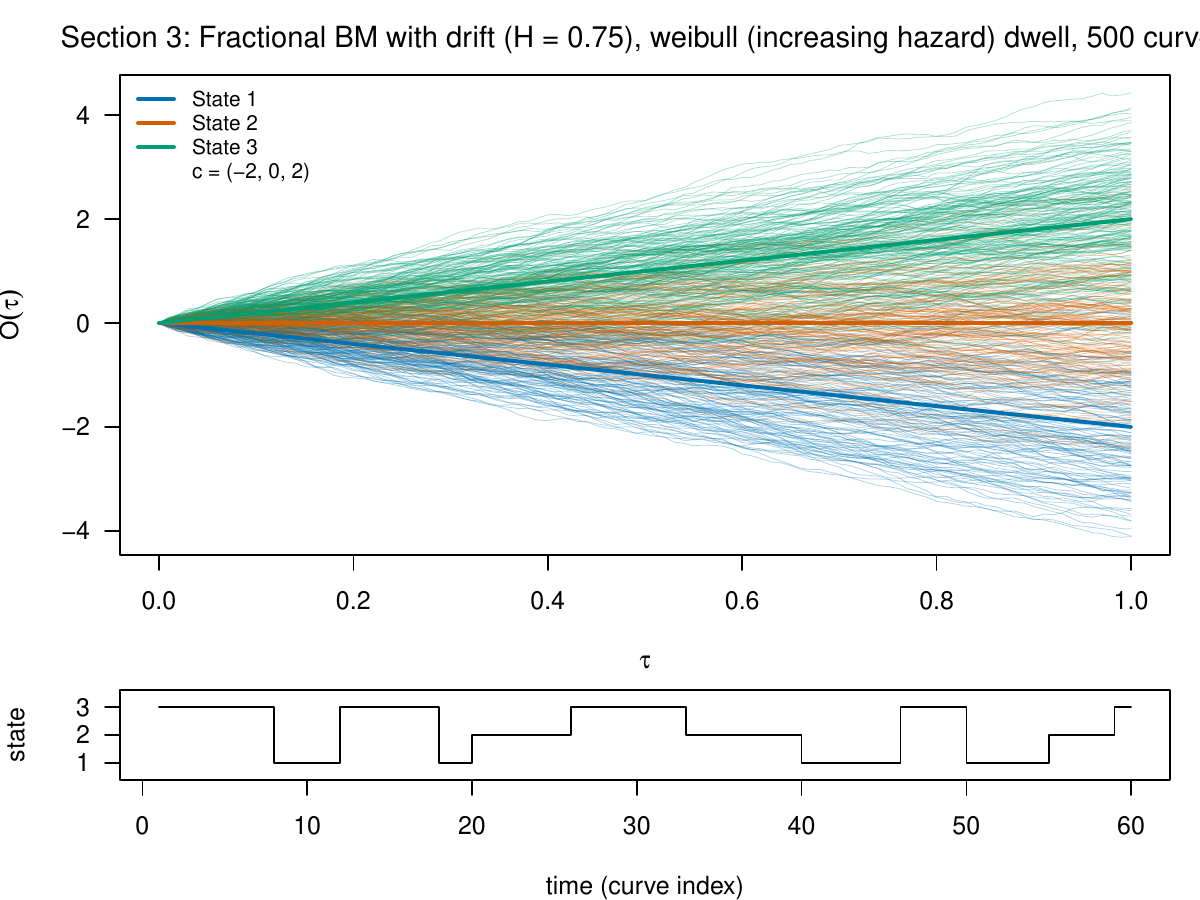}
    \caption{Simulated Fractional BM with drift from a three state semi-Markov process where sojourn distributions are Weibull.}
    \label{fig:fbmW}
\end{figure}

\section{Pediatric Obstructive Sleep Apnea Details}\label{apx:osa}

\begin{figure}[H]
    \centering
    \includegraphics[width=0.8\linewidth]{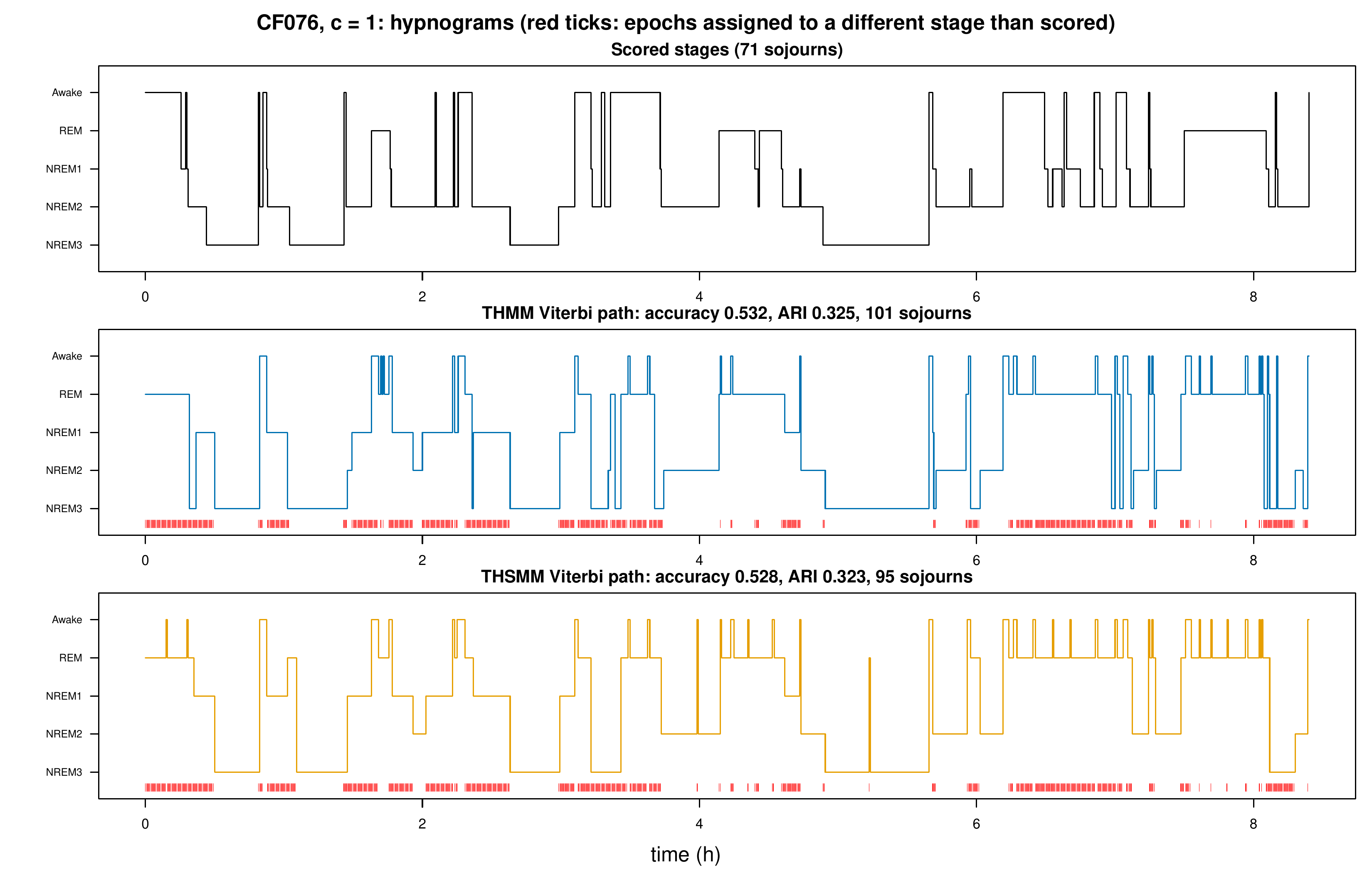}
    \caption{Decoded state sequence from Viterbi algorithm for the THMM and THSMM when
$c = 1.0$.}
    \label{fig:c1viterbi}
\end{figure}

\begin{figure}[H]
    \centering
    \includegraphics[width=0.8\linewidth]{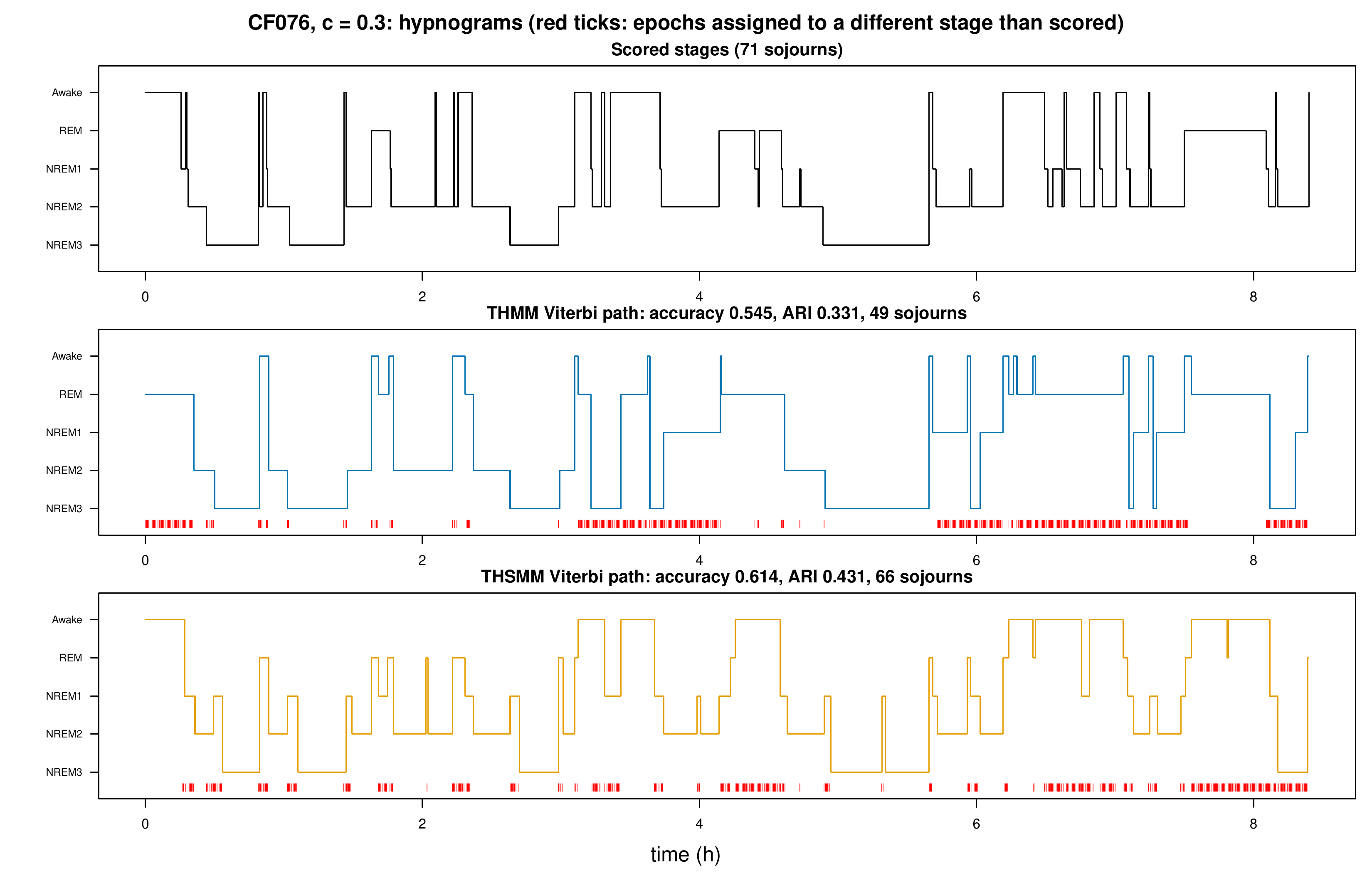}
    \caption{Decoded state sequence from Viterbi algorithm for the THMM and THSMM when $c=0.3$.}
    \label{fig:c03viterbi}
\end{figure}

\begin{figure}[H]
    \centering
    \includegraphics[width=0.8\linewidth]{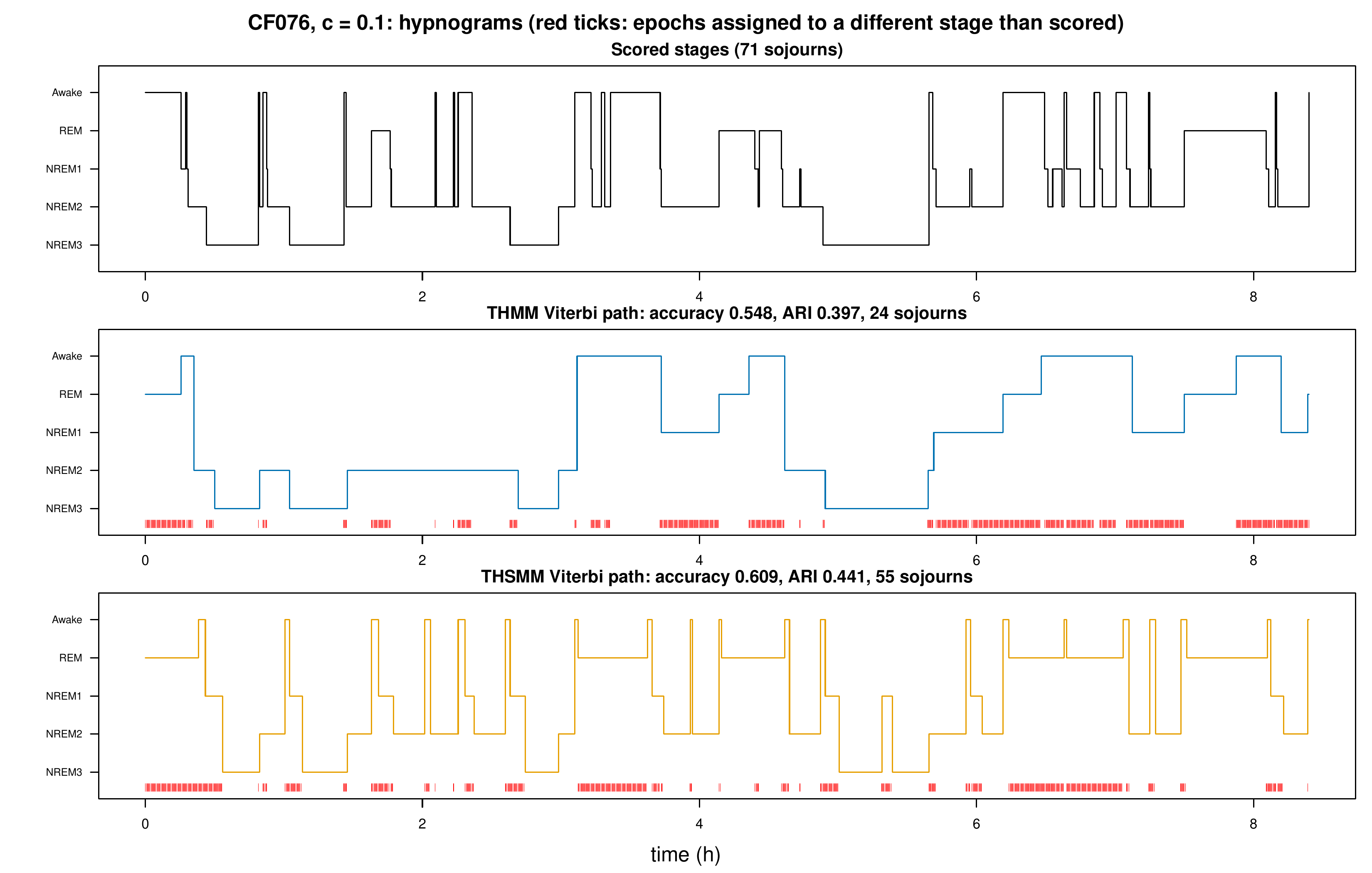}
    \caption{Decoded state sequence from Viterbi algorithm for the THMM and THSMM when
$c = 0.1$.}
    \label{fig:c01viterbi}
\end{figure}

\end{appendices}

\end{document}